\documentclass{article}

\usepackage{microtype}
\usepackage{subfigure}
\usepackage{booktabs} 
\usepackage[table]{xcolor}

\usepackage{hyperref}

\usepackage[accepted]{icml2024}

\usepackage{amsmath}
\usepackage{amssymb}
\usepackage{mathtools}
\usepackage{amsthm}
\usepackage{multirow}

\usepackage[capitalize,noabbrev]{cleveref}

\theoremstyle{plain}

\newtheorem*{theorem*}{Theorem}

\theoremstyle{definition}

\theoremstyle{remark}

\usepackage[textsize=tiny]{todonotes}

\icmltitlerunning{Effective Protein-Protein Interaction Exploration with PPIretrieval}

\begin{document}

\twocolumn[
\icmltitle{Effective Protein-Protein Interaction Exploration with PPIretrieval}

\icmlsetsymbol{equal}{*}

\begin{icmlauthorlist}
\icmlauthor{Chenqing Hua}{mcgill,mila}
\icmlauthor{Connor Coley}{mit}
\icmlauthor{Guy Wolf}{mila,udem}
\icmlauthor{Doina Precup}{mcgill,mila,deepmind}
\icmlauthor{Shuangjia Zheng}{sjtu,aureka}
\end{icmlauthorlist}

\icmlaffiliation{mcgill}{McGill University}
\icmlaffiliation{mila}{Mila-Quebec AI Institute}
\icmlaffiliation{udem}{Université de Montréal}
\icmlaffiliation{mit}{Massachusetts Institute of Technology}
\icmlaffiliation{deepmind}{DeepMind}
\icmlaffiliation{sjtu}{Shanghai Jiao Tong University}
\icmlaffiliation{aureka}{Aureka Biotechnologies}

\icmlcorrespondingauthor{Chenqing Hua}{chenqing.hua@mail.mcgill.ca}
\icmlcorrespondingauthor{Shuangjia Zheng}{shuangjia.zheng@sjtu.edu.cn}

\icmlkeywords{Machine Learning, ICML}

\vskip 0.3in
]



\printAffiliationsAndNotice{}  

\begin{abstract}
Protein-protein interactions (PPIs) are crucial in regulating numerous cellular functions, including signal transduction, transportation, and immune defense. As the accuracy of multi-chain protein complex structure prediction improves, the challenge has shifted towards effectively navigating the vast complex universe to identify potential PPIs. Herein, we propose PPIretrieval, the first deep learning-based model for protein-protein interaction exploration, which leverages existing PPI data to effectively search for potential PPIs in an embedding space, capturing rich geometric and chemical information of protein surfaces.
When provided with an unseen query protein with its associated binding site, PPIretrieval effectively identifies a potential binding partner along with its corresponding binding site in an embedding space, facilitating the formation of protein-protein complexes.
Our codes are available on \url{https://github.com/WillHua127/ppi_search}.
\end{abstract}

\vspace{-0.5cm}
\section{Introduction}
\vspace{-0.2cm}
Proteins are the building blocks of life, engaged in a myriad of interactions \citep{whitford2013proteins}. Understanding how proteins interact with each other is fundamental to unraveling the intricate machinery of biological systems \citep{de2010protein, whitford2013proteins}. Therefore, the ability to predict and analyze protein-protein interactions (PPIs) not only improves our understanding of cellular functions but also plays a pivotal role in drug discovery \citep{fry2006protein, scott2016small, athanasios2017protein}. Current deep learning methods try to analyze, understand, and design PPIs \citep{sun2017sequence, gainza2020deciphering, reau2023deeprank, gao2023hierarchical}, but the results are often constrained by the complexity of protein interactions and the limited understanding of the underlying mechanisms. Despite the progress, there is a pressing need for more effective strategies aimed at designing new protein binders. This objective is crucial for advancing therapeutic interventions and understanding the protein interactions of various biological processes.

Geometric deep learning emerges as a potent strategy for representing and learning about proteins, focusing on structured, non-Euclidean data like graphs and meshes \citep{satorras2021n, luan2020complete, luan2021heterophily,luan2022revisiting, hua2022high, hua2023mudiff}. In this context, proteins can be effectively modeled as graphs, where nodes correspond to individual atoms or residues, and edges represent interactions between them \citep{gligorijevic2021structure, zhang2022protein}. In addition, protein structures can be represented as point clouds or meshes, wherein each point or vertex corresponds to an atom or a residue \citep{gainza2020deciphering, sverrisson2021fast}. Indeed, representing proteins as graphs or point clouds offers a valuable approach for gaining insights into and learning the fundamental geometric and chemical mechanisms governing PPIs \citep{reau2023deeprank, gao2023hierarchical, wang2023learning}. This representation allows for a more comprehensive exploration of the intricate relationships and structural features within protein structures \citep{tubiana2022scannet, isert2023structure}.

Instead of directly generating PPIs, given the limited understanding of their underlying biological mechanisms, we draw inspiration from a previously developed model known as Foldseek \citep{van2023fast}. This model aligns the structure of a query protein against a database by representing its tertiary structure in an embedding space. Rather than attempting to build a generative model that may inaccurately interpret the complex mechanisms of PPIs, we can utilize existing PPI data to develop a retrieval model for PPIs in an embedding space.

Therefore, we introduce the first deep learning-based PPI retrieval model, namely \textbf{PPIretrieval}, which leverages existing PPI data to  search for potential PPIs in an embedding space with rich geometric and chemical information. 
In our approach, PPIretrieval learns surface representations of PPI complexes, which are initially represented as two point clouds along with their corresponding binding interfaces. Then, the embedded surface representations with information about binding partners are stored in our database for further comparison. When provided with an unseen query protein, PPIretrieval learns its surface representation and retrieves the most similar surface representation along with its known binding partner in our database. PPIretrieval outputs the binding partner along with predicted binding interface for the query protein, enabling the exploration of potential PPIs.
PPIretrieval operates solely within a deep learning pipeline, elimitating the need for precomputation of protein patches. This characteristic ensures that PPIretrieval is a fast-computing retrieval tool for efficient and effective PPI exploration.

Overall, a visualization of training and inference workflows of PPIretrieval is illustrated in Fig.~\ref{fig:ppisearch.workflow}. For a thorough understanding of the model architecture, detailed discussions are provided in Sec.~\ref{sec:PPIsearch}. Moreover, we explore the optimization strategy employed by PPIretrieval to learn the \textit{lock-and-key} structure of PPI complexes, which is discussed in Sec.~\ref{sec:training.objective}. We visualize the retrieval results in Sec.~\ref{sec:search.visualization}, and empirically evaluate results in Sec.~\ref{sec:experiment} on real-world PPI datasets. 
Experimental results demonstrate the effectiveness of PPI exploration using PPIretrieval.
Most importantly, we discuss future work and directions in Sec.~\ref{sec:future.work}.

\vspace{-0.3cm}
\section{Preliminaries}
\vspace{-0.2cm}
In the context of PPIretrieval, the objective is to identify a binding partner $B$ with its corresponding binding interface for an unseen query protein $P$ with a known binding interface. The approach involves learning protein interactions through their surface manifolds. Specifically, the model employs heat diffusion on the surface, learning individual representations. The goal is to effectively retrieve potential binding partners and their associated binding interfaces based on the learned surface representations.

\noindent \textbf{Protein Surface Representation} \quad
A protein is a chain of residues ${P}=\{\mathbf{a}_1,...,\mathbf{a}_{N_P}\}\in [0,1]^{N_P\times 20}$, consisting of $N_P$ residues in Euclidean space $\upsilon_P=\{r_1, ..., r_N\}\in \mathbb{R}^{N_P\times 3}$. The binding interface $\mathbf{Y}^{\text{res}}_{{P}} \in \{0, 1\}^{N_P\times 1}$ of protein $P$ denotes a region where the protein is poised to interact with another protein, forming a complex. The protein $P$ can be characterized by a set of surface points $\{x_1,...,x_{M_P}\}\in \mathbb{R}^{M_P\times 3}$ with unit normals $\{n_1,...,n_{M_P}\}\in \mathbb{R}^{M_P\times 3}$ located on its surface. In this context, we employ the surface representation to learn the protein, which captures its structural characteristics and potential interactions with other proteins.

\noindent \textbf{LB Operator on Protein Surface} \quad
The Laplace-Beltrami (LB) operator is known for its smooth operation on compact Riemannian manifolds \citep{levy2006laplace, wang2023learning}.
When applied to the protein surface, which can be considered as a 2D Riemannian manifold $\mathcal{M}$ with a Laplace-Beltrami (LB) operator $\Delta_\mathcal{M}$, the LB operator operates smoothly.
The LB operator $\Delta$ has an eigendecomposition $\Delta \phi_i = \lambda_i \phi_i, \ 0\leq \lambda_1 \leq \lambda_2 \leq ...$, and the set $\{\phi_1,\phi_2,...\}$ forms orthonormal basis for the space of functions defined on $\mathcal{M}$. Any function on the surface can be expressed as a linear combination of these eigenfunctions,
\begin{equation}
\vspace{-0.1cm}
\label{eq:basis.function}
\resizebox{0.5\hsize}{!}{$
\small
    g = \sum_i a_i \phi_i,\ a_i = \langle g,\phi_i \rangle_{\mathcal{M}}.
$}
\end{equation}
Here, $g$ is a linear combination of basis functions, with the scalars $a_i$ determined by the inner product on the surface.

\noindent \textbf{LB Operator-Induced Message-passing} \quad
Consider $g(x,t)$ as the measure of heat at point $x$ on the surface at time $t$. The heat operation is a message-passing process on the surface, propagating from hot regions to cool regions. The change in heat over time is described by the LB operator $\frac{\partial g}{\partial t} = \Delta g$ \citep{levy2006laplace}. After time $t$, the heat distribution is equivariant to
\vspace{-0.2cm}
\begin{equation}
\vspace{-0.2cm}
\resizebox{0.7\hsize}{!}{$
\small
    g(x, t) = \exp{(H_t)}g,\ H_t = -\Delta t = -\lambda t.
$}
\end{equation}
where $H_t$ denotes the heat operator at time $t$. Following Eq.~\ref{eq:basis.function}, we can define a function propagation operator $\mathcal{F}$ on the heat distribution over the surface \citep{wang2023learning},
\begin{equation}
\label{eq:heat.operator}
\vspace{-0.2cm}
\resizebox{0.7\hsize}{!}{$
\small
    \mathcal{F}g(x,t) = \sum_i F_i \exp(-\lambda_i t) \langle g,\phi_i \rangle_{\mathcal{M}} \phi_i,
$}
\end{equation}
where $\{F_i\}$ are frequency filters, $\{\lambda_i\}$ are eigenvalues, $\{\phi_i\}$ are eigenfunctions, and $t$ is a time parameter for the operator.
\citep{wang2023learning} propose to use a linear Gaussian filter to learn frequency, $F_i=F_{(\mu,\sigma)}(\lambda_i) = \exp (-\frac{(\lambda_i-\mu)^2}{\sigma^2})$, where $\mu, \sigma^2$ are learned parameters for mean and variance. 

This LB operator-induced message-passing is employed for smooth and effective information flow on the (compact) protein surface, as discussed in Sec.~\ref{sec:surface.encoder}.

\vspace{-0.3cm}
\section{PPIretrieval}
\vspace{-0.2cm}
\begin{figure*}[ht!]
\centering
{
\includegraphics[width=1\textwidth]{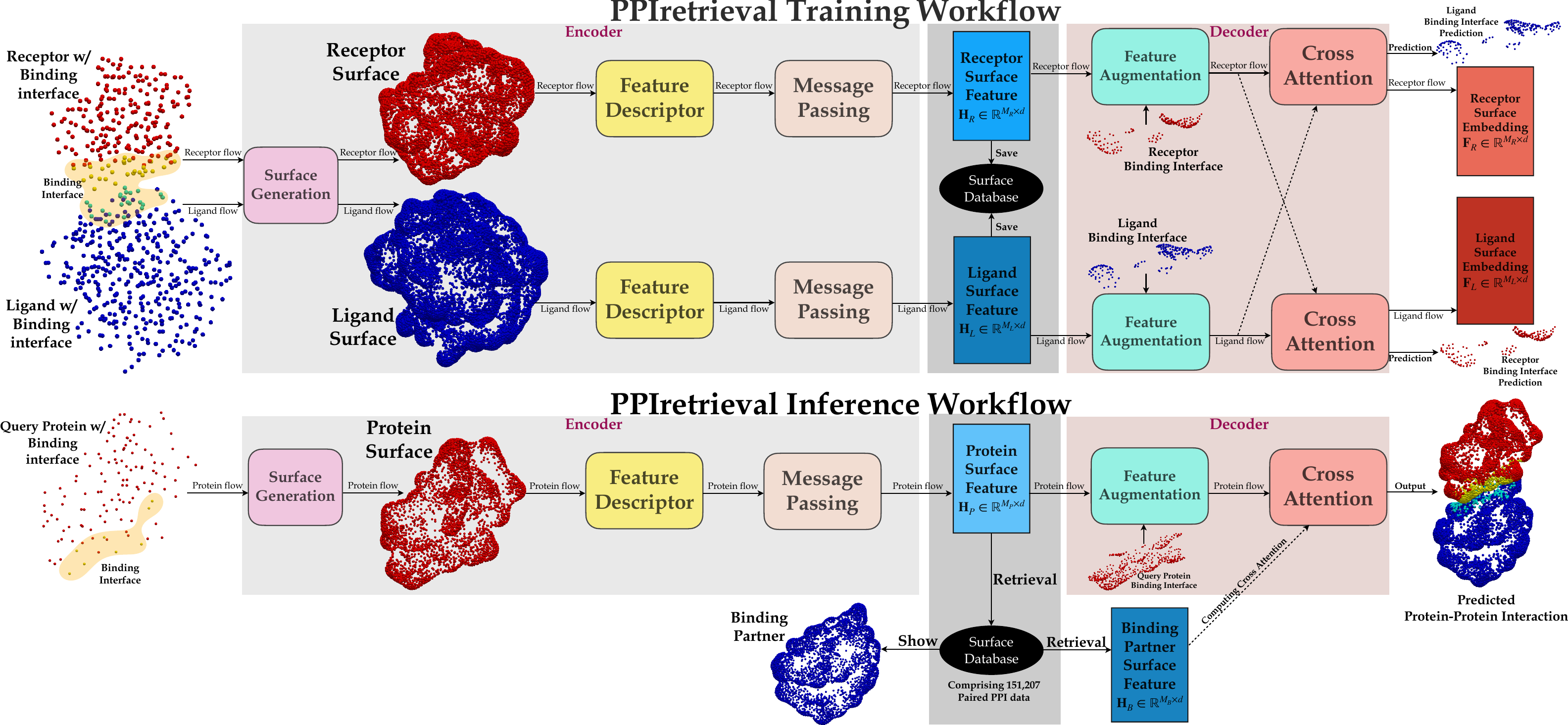}}
\vspace{-0.8cm}
{
  \caption{
  An overview of PPIretrieval pipeline, demonstrating the training and inference workflows. \textbf{During training}, PPIretrieval processes a PPI complex. The encoder network (shown in grey) encodes the two proteins, generating surface features, $\mathbf{H}_R,\mathbf{H}_L$. These features, along with information about their binding partnership, are stored in our database. The decoder network (shown in brown) then takes these surface features, along with the receptor's binding interface as input, predicting the ligand's binding interface and generating its embedding $\mathbf{F}_L$, and vice versa.
  \textbf{During inference}, PPIretrieval takes a protein $P$ with its corresponding binding interface to the encoder network. It encodes $P$ into a surface feature $\mathbf{H}_P$. Then, PPIretrieval identifies a surface feature $\mathbf{H}_B$ for binding partner $B$ in our database. The decoder network takes the surface features $\mathbf{H}_P, \mathbf{H}_B$, along with $P$'s binding interface, predicting $B$'s binding interface. A PPI complex is predicted between the input protein $P$ with the given binding interface and the predicted binding partner $B$ with the predicted binding interface, as demonstrated above. Details of PPIretrieval can be found in Sec.~\ref{sec:PPIsearch}, with an overview discussed in Sec.~\ref{sec:ppisearch.pipeline.overview}.
  }
  \label{fig:ppisearch.workflow}
  \vspace{-0.5cm}
}
\end{figure*}

\label{sec:PPIsearch}
The overview of PPIretrieval is demonstrated in Fig.~\ref{fig:ppisearch.workflow}.
PPIretrieval follows a specific design process. Starting with a query protein ${P}$ comprising $N_P$ residues in Euclidean space $\upsilon_P \in \mathbb{R}^{N_P\times 3}$,  along with its corresponding binding interface $\mathbf{Y}^{\text{res}}_{{P}}$, we first encode the protein into a surface representation $\textbf{H}_{{P}} \in \mathbb{R}^{M_{{P}} \times d}$. The next step involves retrieving the most similar surface feature $\textbf{H}_{A} \in \mathbb{R}^{M_{{A}} \times d}$ in the database. Once identified, we locate a binding partner protein ${B}$ that binds to ${A}$, with a surface feature $\mathbf{H}_{B} \in \mathbb{R}^{M_{{B}} \times d}$. Then, the model decodes the pair of protein surface features, $\textbf{H}_{{P}}, \textbf{H}_{{B}}$, utilizing the known binding interface $\mathbf{Y}^{\text{res}}_{{P}}$. The final prediction involves estimating the binding interface $\hat{\mathbf{Y}}^{\text{res}}_{{B}}$ for protein ${B}$. In summary, the model outputs the protein ${B}$ that is most likely to bind to the input protein ${P}$, accompanied by the predicted binding interface $\hat{\mathbf{Y}}^{\text{res}}_{{B}}$ .

It is important to highlight that our entire pipeline operates on a deep-learning framework. There is no need for precomputation of protein or surface patches, making PPIretrieval an efficient tool. This characteristic enables a fast encoding for the input protein and facilitates effective retrieval for the output protein within the model.

\vspace{-0.2cm}
\subsection{Surface Representation Encoder}
\vspace{-0.2cm}
\label{sec:surface.encoder}
The surface encoder network aims to encode an input protein ${P}$ into a surface representation $\mathbf{H}_{P} \in \mathbb{R}^{M_{P} \times d}$, where $M_P$ denotes the number of surface points representing $P$. This representation captures the propagated chemical and geometric information. It can be stored in our database for subsequent retrieval and comparison purposes.

\noindent \textbf{Protein Surface Preparation} \quad
The input consists of one-hot encoded residue types for protein ${P} \in [0,1]^{N_P\times 20}$ in Euclidean space $\upsilon_P \in \mathbb{R}^{N_P\times 3}$, along with the corresponding binding interface $\mathbf{Y}^{\text{res}}_{P} \in \{0, 1\}^{N_P\times 1}$. We apply dMaSIF \citep{sverrisson2021fast} to generate a set of oriented surface points $\{x_1,...,x_{M_P}\}\in \mathbb{R}^{M_P\times 3}$ with unit normals $\{n_1,...,n_{M_P}\}\in \mathbb{R}^{M_P\times 3}$ to approximate a smooth manifold representing the surface of protein ${P}$.

Upon approximating the protein surface with surface points, we define the LB operator $\Delta_{P}$ on the surface for heat diffusion.
We compute the first $k$ eigenfunctions of $\Delta_{P}$ stacked in matrix $\Phi_{P} \in\mathbb{R}^{M\times k}$ with their corresponding eigenvalues $\{\lambda_i\}_{i=1}^k$ ($k=100$), then calculate the Moore-Penrose pseudo-inverse of the eigenfunction matrix, $\Phi^+_{P} \in\mathbb{R}^{k\times M}$. The heat diffusion allows effective information flow between surface points, crucial for operating heat diffusion on the surface \citep{wang2023learning}.

\noindent \textbf{Geometric Descriptor} \quad
Following the computation of surface points to represent the input protein, we describe the local geometric features of the surface.
We approximate the per-point mean curvature, Gaussian curvature as detailed in \citep{cao2019efficient}, and compute the Heat Kernel Signatures as described in \citep{sun2009concise}. The geometric features processed by a MLP, $\mathbf{F}_{\text{Geom}} \leftarrow \text{MLP}([\mathbf{F}_{\text{Mean}}, \mathbf{F}_{\text{Gauss}}, \mathbf{F}_{\text{HKS}}]) \in \mathbb{R}^{M_P\times d_G}$, capture the local geometric environment for each surface point. 

\noindent \textbf{Chemical Descriptor} \quad
We proceed to compute chemical features for surface points based on the one-hot encoded residue types and the binding interface. The residue-level chemical features are first encoded using a MLP with concatenated features, $\mathbf{F}_{\text{Res}} \leftarrow \text{MLP}([{P}, \mathbf{Y}^{\text{res}}_P]) \in \mathbb{R}^{N_P \times d_C}$, then transformed by an equivariant-GNN \citep{satorras2021n}, $\mathbf{F}_{\text{Res}} \leftarrow \text{EGNN}(\mathbf{F}_{\text{Res}}, \upsilon_P) \in \mathbb{R}^{N_P \times d_C}$. The residue-level features $\mathbf{F}_{\text{Res}}$ preserve the local information pertaining to residues and the binding interface.

Then, we project these residue-level features onto surface-level features. For each surface point $x_i$,  we identify its $k$ nearest neighboring residues $\{r^i_1,...,r^i_k\}$ with features $\{\mathbf{f}^{i,1}_{\text{Res}},..., \mathbf{f}^{i,k}_{\text{Res}}\}$. A vector of chemical features is computed by applying a MLP and summation over the nearest neighboring residues with a distance filter\footnote{We use a cosine cutoff function, $f_{\cos}(d) = \frac{1}{2}(\cos(\frac{\pi d}{d_\text{cut}}) +1)$, to smooth out the distance transition to $0$ as the distance $d$ approaches a pre-defined cutoff distance $d_\text{cut}=30\mathring{\text{A}}$.}, $\mathbf{f}^i_\text{Chem} \leftarrow \text{MLP} (\sum_{j=1}^k f_{\cos}(\|x_i - r^i_j\|) \cdot \text{MLP}([\mathbf{f}^{i, j}_\text{Res}, 1/\|x_i - r^i_j\|])) \in \mathbb{R}^{d_C}.$ By computing and stacking these per-point features, we obtain the chemical features for the protein surface $\mathbf{F}_\text{Chem}=\{\mathbf{f}^1_\text{Chem},..., \mathbf{f}^M_\text{Chem}\} \in \mathbb{R}^{M_P\times d_C}$.

Finally, we use a MLP to combine the per-point geometric and chemical features, $\mathbf{F}_\text{Surf} \leftarrow \text{MLP}([\mathbf{F}_\text{Geom}, \mathbf{F}_\text{Chem}]) \in \mathbb{R}^{M_P\times d}$. For each surface point, these features capture the local geometric and chemical environment, along with binding interface information. These can be effectively used in the heat diffusion process for facilitating information flow on the approximated protein surface.

\noindent \textbf{Message-passing on Protein Surface} \quad
We perform heat operation on protein surfaces for message passing. This operation can be smoothly applied to the protein surface, treating it as a compact object \citep{wang2023learning}.
For a protein with surface feature $\mathbf{F}_{\text{Surf}}$, we first project it onto the column space of eigenfunctions, $\mathbf{F}'_{\text{Surf}} \leftarrow \Phi^+_{{P}} \mathbf{F}_\text{Surf} \in \mathbb{R}^{k\times d}$, expressing features in the orthogonal basis with a reduced dimension. Then following Eq.~\ref{eq:heat.operator}, we can perform heat operation on the surface as,
\vspace{-0.2cm}
\begin{equation}
\vspace{-0.2cm}
\resizebox{0.91\hsize}{!}{
\small
    $\mathbf{H}_{P} = \Phi 
\underbrace{
\begin{pmatrix}
e^{-\frac{(\lambda_1 - \mu_1)^2}{\sigma_1^2} -\lambda_1 t_1} & \cdots & e^{-\frac{(\lambda_1 - \mu_d)^2}{\sigma_d^2} -\lambda_1 t_d}\\
\vdots  & \ddots & \vdots  \\
e^{-\frac{(\lambda_k - \mu_1)^2}{\sigma_1^2} -\lambda_k t_1} & \cdots & e^{-\frac{(\lambda_k - \mu_d)^2}{\sigma_d^2} -\lambda_k t_d}
\end{pmatrix} 
}_{=F \exp(-\lambda t) \in \mathbb{R}^{k \times d}}
\odot \mathbf{F}'_{\text{Surf}} \in \mathbb{R}^{M\times d}.$
}
\end{equation}
Here, each feature channel of $\mathbf{F}_\text{Surf}$ has its unique set of $\{\mu_i, \sigma_i, t_i\}_{i=1}^d$. In summary, the encoder network encodes a protein (represented by residues) into a surface representation by applying heat diffusion on the surface, capturing both local geometric and chemical environments.

\noindent \textbf{Training Stage} \quad
During training, the encoder network is fed with the paired receptor and ligand proteins $R, L$, along with their corresponding binding interfaces $\mathbf{Y}^{\text{res}}_R, \mathbf{Y}^{\text{res}}_L$. It then processes these inputs to generate distinct surface features for each protein,  
\begin{equation}
\vspace{-0.2cm}
\resizebox{0.9\hsize}{!}{
\small
$
    \mathbf{H}_R = \text{Enc}(R, \mathbf{Y}^{\text{res}}_R) \in \mathbb{R}^{M_R\times d},\ \mathbf{H}_L = \text{Enc}(L, \mathbf{Y}^{\text{res}}_L) \in \mathbb{R}^{M_L\times d}.
$
    }
\end{equation}

\noindent \textbf{Binding Interface of Surface} \quad
We construct the surface-level binding interface $\mathbf{Y}^{\text{surf}}_P \in \{0,1\}^{M_P \times 1}$ for surface points based on the residue-level binding interface $\mathbf{Y}^{\text{res}}_P \in \{0,1\}^{N_P \times 1}$ for protein $P$. Here,  $\mathbf{y}^{\text{surf}}_{i,P}=1$ indicates that the surface point $x_i$ belongs to the binding interface. 
For each residue in $P$, we define a region with a predefined radius of $r=10\mathring{\text{A}}$. All surface points falling within this region are then labeled as part of the surface binding interface.



\vspace{-0.2cm}
\subsection{Interactive Decoder}
\vspace{-0.2cm}
\label{sec:interactive.decoder}
The decoder network operates by taking surface features as input, allowing interaction between two proteins, and ultimately predicting a binding interface.

We assume that input data comprises the receptor $R$ and its corresponding binding interface $\mathbf{Y}^{\text{res}}_R$. The encoder network generates a surface feature $\mathbf{H}_R \in \mathbb{R}^{M_R\times d}$. Simultaneously, we compute the surface-level binding interface $\mathbf{Y}^{\text{surf}}_R \in \{0,1\}^{M_R\times 1}$. Then, the model identifies a binding partner $L$ and obtains its own surface feature $\mathbf{H}_L \in \mathbb{R}^{M_L\times d}$. The objective is to predict the binding interface for ligand $L$, expressed as $p(\hat{\mathbf{Y}}^{\text{res}}_L|\mathbf{H}_R, \mathbf{H}_L, \mathbf{Y}^{\text{surf}}_R) \in [0,1]^{N_L\times 1}$.

\noindent \textbf{Cross-Attention} \quad
Before computing the cross-attention, we update the surface features for receptor and ligand using an equivariant-GNN and MLPs, $\mathbf{H}_R \leftarrow \text{EGNN}(\text{MLP}([\mathbf{H}_R, \mathbf{Y}^{\text{surf}}_R]), x_R) \in \mathbb{R}^{M_R \times d}, \mathbf{H}_L \leftarrow \text{EGNN}(\text{MLP}(\mathbf{H}_L), x_L) \in \mathbb{R}^{M_L \times d}$. Here, $\mathbf{H}_R$ is updated with information about the binding interface, providing the model with improved capabilities for locating the binding interface of the ligand $L$ during cross-attention.

Given the updated receptor features $\mathbf{H}_R$ and ligand features $\mathbf{H}_L$, we compute the cross-attention between two protein surface features, enabling interaction and communication,
\begin{align}
\vspace{-0.2cm}
    &\scalebox{0.8}{$
    \begin{aligned}
        \mathbf{F}_R &= \text{softmax}\left(\frac{(\mathbf{H}_R W_{\text{Q}})(\mathbf{H}_L W_{\text{K}})^T}{\sqrt{d}}\right) (\mathbf{H}_L W_{\text{V}}), \nonumber\\
        \mathbf{F}_L &= \text{softmax}\left(\frac{(\mathbf{H}_L W_{\text{Q}})(\mathbf{H}_R W_{\text{K}})^T}{\sqrt{d}}\right) (\mathbf{H}_R W_{\text{V}}),
    \end{aligned}$}
\end{align}
where $W_{\text{Q}}, W_{\text{K}}, W_{\text{V}} \in \mathbb{R}^{d\times d}$ are learned parameters for the query, key, value in the attention mechanism, respectively. This facilitates effective interaction and communication between the receptor and ligand surface features.

\noindent \textbf{Binding Interface} \quad
Once we obtain the propagated surface features for the ligand $\mathbf{F}_L\in \mathbb{R}^{M_L\times d}$, we employ a MLP with sigmoid function directly on these features, predicting for the binding interface, $\hat{\mathbf{Y}}^{\text{surf}}_L \leftarrow \sigma(\text{MLP}(\mathbf{F}_L))\in [0,1]^{M_L\times 1}$. Thus, the prediction of surface-level binding interface for ligand $L$ is solely conditioned by the surface features of both the receptor and the ligand, along with the binding interface information of the receptor.

\noindent \textbf{Surface Point to Residue} \quad
We compute the residue-level binding interface $\hat{\mathbf{Y}}^{\text{res}}_L \in \{0, 1\}^{N_L \times 1}$ from $\hat{\mathbf{Y}}^{\text{surf}}_L$, and embedding $\hat{\mathbf{F}}_L \in \mathbb{R}^{N_L \times d}$ from $\mathbf{F}_L$. For each residue $i$ in $L$, we define a region with a fixed radius of $r=10\mathring{\text{A}}$ and collect a set of surface points within this region, each with a binding interface $\hat{\mathbf{y}}^{\text{surf}}_j$ and embedding $\mathbf{f}_j$. The residue $i$ is considered part of the binding interface if the majority of surface points in the region are labeled as part of the binding interface, i.e., $\hat{\mathbf{y}}^{\text{res}}_i = 1$ if $\text{Mean}(\sum_j \hat{\mathbf{y}}^{\text{surf}}_j) > 0.5$; otherwise $\hat{\mathbf{y}}^{\text{res}}_i = 0$. And the residue-level embedding is obtained using the same logic, where $\hat{\mathbf{f}}_i = \text{Mean}(\sum_j {\mathbf{f}}_j)$.

\noindent \textbf{Training Stage} \quad
During training, each PPI sample is treated as two training instances. The model first takes the receptor $R$ and its associated binding interface $\mathbf{Y}^{\text{surf}}_R$ as input, predicting the ligand's binding interface; then the model takes the ligand $L$ and its corresponding binding interface $\mathbf{Y}^{\text{surf}}_L$ as input, predicting the receptor's binding interface,
\begin{equation}
\vspace{-0.2cm}
\resizebox{0.9\hsize}{!}{
$
\small
    \hat{\mathbf{Y}}^{\text{res}}_L, \hat{\mathbf{F}}_L  = \text{Dec}(\mathbf{H}_R, \mathbf{H}_L, \mathbf{Y}^{\text{surf}}_R),\  \hat{\mathbf{Y}}^{\text{res}}_R, \hat{\mathbf{F}}_R = \text{Dec}(\mathbf{H}_R, \mathbf{H}_L, \mathbf{Y}^{\text{surf}}_L).
$
}
\end{equation}

\vspace{-0.2cm}
\subsection{Overview: PPIretrieval Pipeline}
\vspace{-0.2cm}
\label{sec:ppisearch.pipeline.overview}

\noindent \textbf{Training Stage} \quad
During training, PPIretrieval processes a PPI complex with their corresponding binding interface. The encoder network encodes the two proteins, resulting in two surface features $\mathbf{H}_R, \mathbf{H}_L$, respectively. These surface representations, along with information about their binding partnership, are stored in our database. Then, the decoder network takes the surface features, $\mathbf{H}_R, \mathbf{H}_L$, along with the receptor's binding interface $\mathbf{Y}^{\text{surf}}_R$ as input, and predicts the ligand's binding interface $\hat{\mathbf{Y}}^{\text{res}}_L$. This process is repeated to predict the receptor's binding interface $\hat{\mathbf{Y}}^{\text{res}}_R$ vice versa. PPIretrieval undergoes optimization to utilize the \textit{lock-and-key} structures of the PPI complex, following a specific approach discussed in Sec.~\ref{sec:training.objective}. This training setup ensures that PPIretrieval learns to predict binding interfaces for both proteins involved in a PPI complex.

\noindent \textbf{Inference Stage} \quad
During inference, PPIretrieval takes an unseen query protein $P$ along with a specified binding interface to the encoder network. It encodes $P$ into a surface feature $\mathbf{H}_P$. Then, PPIretrieval searches our database to retrieve the most similar surface feature $\mathbf{H}_A$ using a similarity function. Once the match is found, the surface feature $\mathbf{H}_B$ of the binding partner $B$ that binds to $A$ is identified. Then, the decoder network takes the surface features $\mathbf{H}_P, \mathbf{H}_B$, along with $P$'s given binding interface $\mathbf{Y}^{\text{surf}}_P$, and predicts $B$'s binding interface, $\hat{\mathbf{Y}}^{\text{res}}_B$.
This process entails the model making predictions on how the surface of the binding partner interacts with the provided protein and its binding interface. Finally, PPIretrieval outputs protein $B$ along with the predicted binding interface, indicating the most likely binding scenario with protein $P$ and its given binding interface. A visual demonstration of this process is illustrated in Fig.~\ref{fig:ppisearch.workflow}.

\vspace{-0.3cm}
\section{Training Objective}
\vspace{-0.2cm}
\label{sec:training.objective}
Consider $\hat{\mathbf{F}}_{R} \in \mathbb{R}^{N_{R} \times d}, \hat{\mathbf{F}}_{L} \in \mathbb{R}^{N_{L} \times d}$ as the propagated surface features derived from our interactive decoder model for the receptor and ligand proteins.
Additionally, let $\hat{\mathbf{Y}}^{\text{res}}_{R} \in [0,1]^{N_{R} \times 1}, \hat{\mathbf{Y}}^{\text{res}}_{L} \in [0,1]^{N_{L} \times 1}$ denote the predicted binding interface for the receptor and ligand protein, respectively. The optimization aims to utilize the \textit{lock-and-key} structure between the receptor and ligand in a PPI complex.

\vspace{-0.2cm}
\subsection{Lock-and-Key Optimization}
\vspace{-0.2cm}
In our model, we assume an entirely rigid protein structure. Within the protein complex, a \textit{lock-and-key} structure is established between the rigid proteins, where their structures exhibit complementary representations \citep{morrison2006lock}.
To utilize the structure, we optimize the model to learn pairwise matching between residue features $\hat{\mathbf{F}}_{R}, \hat{\mathbf{F}}_{L}$, drawing inspiration from shape correspondences \citep{jain2006robust}.
This enables each surface point to identify its \textit{lock-and-key} counterpart in the opposite protein, leveraging the information encoded in the surface features. This unique feature empowers PPIretrieval to effectively assemble two binding sits. Following a similar approach outlined in \citep{lu2023jigsaw} for assembling fractures in vision-related tasks, we first compute an affinity matrix between the surface features of the two proteins.

\noindent \textbf{Affinity Metric} \quad
Given $\hat{\mathbf{F}}_{R}, \hat{\mathbf{F}}_{L}$,
we compute the global affinity matrix $\mathbf{A} \in \mathbb{R}^{N_{R} \times N_{L}}$ and its corresponding doubly-stochastic matrix $\hat{\mathbf{X}}\in [0,1]^{N_R \times N_L}$ as follows,
\begin{equation}
\vspace{-0.2cm}
    \small
    \mathbf{A} = \exp{(\frac{\hat{\mathbf{F}}_{R}^T W \hat{\mathbf{F}}_{L}}{\tau_{\mathbf{A}}})},\ 
    \hat{\mathbf{X}} = \text{sinkhorn}(\mathbf{A}),
\end{equation}
where $W \in \mathbb{R}^{d\times d}$ consists of learnable affinity weights and $\tau_{\mathbf{A}}$ denotes the temperature hyperparameter. $\hat{\mathbf{X}}$ is a doubly-stochastic matrix computed by the differentiable sinkhorn layer \citep{cuturi2013sinkhorn}, where $\hat{\mathbf{X}}_{ij}$ measures the soft-matching score between surface features $\hat{\mathbf{f}}_i \in \hat{\mathbf{F}}_{R}, \hat{\mathbf{f}}_j \in \hat{\mathbf{F}}_{L}$.

\noindent \textbf{Lock-and-Key} \quad
We construct a ground-truth matching matrix $\mathbf{X} \in \{0, 1\}^{N_R \times N_L}$ to optimize the soft-matching score between two proteins as follows,
\begin{equation}
\vspace{-0.2cm}
\resizebox{0.4\hsize}{!}{$
\small
\mathbf{x} _{ij} = 
\begin{cases}
    1, & \text{if } d_{ij} \leq d_{\text{cut}} \\
    0,              & \text{otherwise}.
\end{cases}
$
}
\end{equation}
Here $\mathbf{x}_{ij}=1$ only if the pairwise Euclidean distance between two residues $i\in R, j\in L$ is within a cutoff distance $d_\text{cut}=10\mathring{\text{A}}$, implying that they are close enough to interact; otherwise $\mathbf{x}_{ij}=0$.

To optimize PPIretrieval from the \textit{lock-and-key} perspective, we enforce the soft-matching score to closely resemble the ground-truth matching matrix,
\begin{equation}
\vspace{-0.2cm}
\resizebox{0.91\hsize}{!}{$
\small
    \mathcal{L}_{\text{match}} = -\sum_{1<i<N_R} \sum_{1<j<N_L} \mathbf{x}_{ij} \log \hat{\mathbf{x}}_{ij} + (1-\mathbf{x}_{ij}) \log (1-\hat{\mathbf{x}}_{ij}).
    $}
\end{equation}
The matching loss serves the dual purpose of encouraging a close alignment between the soft-matching scores and the ground truth, as well as providing global rigidity guidance by ensuring each residue is matched with its complementary part in the opposite protein. This approach reinforces the \textit{lock-and-key} structure within the PPI.

\vspace{-0.2cm}
\subsection{Contrastive Optimization}
\vspace{-0.2cm}
In addition to the the \textit{lock-and-key} optimization objective, we aim to bring the residue features of the binding interface closer while pushing residue features that do not belong to the binding interface farther apart. To achieve this, we employ the contrastive loss \citep{xie2020pointcontrast} designed for point clouds. This loss minimizes the distance between the residue features of corresponding residues and maximizes the distance between non-corresponding residues as,
\begin{equation}
\vspace{-0.2cm}
\resizebox{0.8\hsize}{!}{$
\small
    \mathcal{L}_{\text{contra}} = -\sum_{i\in S_R} \sum_{j\in S_L} \log \frac{\exp(\hat{\mathbf{f}}^i_R \cdot \hat{\mathbf{f}}^j_L/\tau_\text{c})}{\sum_{k\in S_L}\exp(\hat{\mathbf{f}}^i_R \cdot \hat{\mathbf{f}}^k_L/\tau_\text{c})}.
$}
\end{equation}
Here $S_R=(\mathbf{Y}^{\text{res}}_R=1), S_L=(\mathbf{Y}^{\text{res}}_L=1)$ are residues of binding interface for receptor and ligand, and $\tau_\text{c}$ is the temperature hyperparameter. 
This objective effectively minimizes the distance between residue features $\hat{\mathbf{f}}^i_{R} \in \hat{\mathbf{F}}_{R}, \hat{\mathbf{f}}^j_{L} \in \hat{\mathbf{F}}_{L}$ of corresponding binding interfaces, enhancing the proximity of relevant residue features.

\vspace{-0.2cm}
\subsection{Binding Interface Optimization}
\vspace{-0.2cm}
\label{sec:binding.intergace.optimization}
In addition to the \textit{lock-and-key} and contrastive optimization objectives, we directly optimize the predictions of binding interfaces, $\hat{\mathbf{Y}}^{\text{res}}_{R}, \hat{\mathbf{Y}}^{\text{res}}_{L}$. Given the ground-truth binding interfaces ${\mathbf{Y}}^{\text{res}}_{R} \in \{0,1\}^{N_R \times 1}, {\mathbf{Y}}^{\text{res}}_{L} \in \{0,1\}^{N_L \times 1}$, we aim to make the predictions close to the ground-truth values as,
\begin{align}
\vspace{-0.3cm}
    &\scalebox{0.8}{
    $
    \begin{aligned}
        \mathcal{L}_{\text{bind}} = &-\sum_{1<i<N_R} \mathbf{y}^{\text{res}}_{i,R} \log \hat{\mathbf{y}}^{\text{res}}_{i, R} + (1-\mathbf{y}^{\text{res}}_{i, R}) \log (1-\hat{\mathbf{y}}^{\text{res}}_{i, R}) \\
    & -\sum_{1<i<N_L} \mathbf{y}^{\text{res}}_{i, L} \log \hat{\mathbf{y}}^{\text{res}}_{i, L} + (1-\mathbf{y}^{\text{res}}_{i, L}) \log (1-\hat{\mathbf{y}}^{\text{res}}_{i, L}).
    \end{aligned}
    $
    }
\end{align}
This objective ensures that the model makes accurate predictions for the binding interfaces of both the receptor and ligand proteins by minimizing the difference between predicted and ground-truth values.

The total loss is the sum of the three loss terms $\mathcal{L} = \mathcal{L}_\text{match} + \mathcal{L}_\text{contra} + \mathcal{L}_\text{bind}$. This optimization is designed to leverage the \textit{lock-and-key} structure inherent in PPI complexes.

\vspace{-0.3cm}
\section{Retrieval Visualization}
\vspace{-0.2cm}
\label{sec:search.visualization}
We visualize some retrieval results, showing the predictions of PPIretrieval when provided with an unseen query protein and observing its potential binding partner within our protein surface database. The database currently comprises a total of $151,207$ paired proteins with $302,414$ surface representations, trained and embedded using PPIs from PDB, DIPS, and PPBS training and validation sets (see Sec.~\ref{sec:experiment}). 
\begin{figure}[ht!]
\vspace{-0.3cm}
\centering
{
\includegraphics[width=1.07\columnwidth]{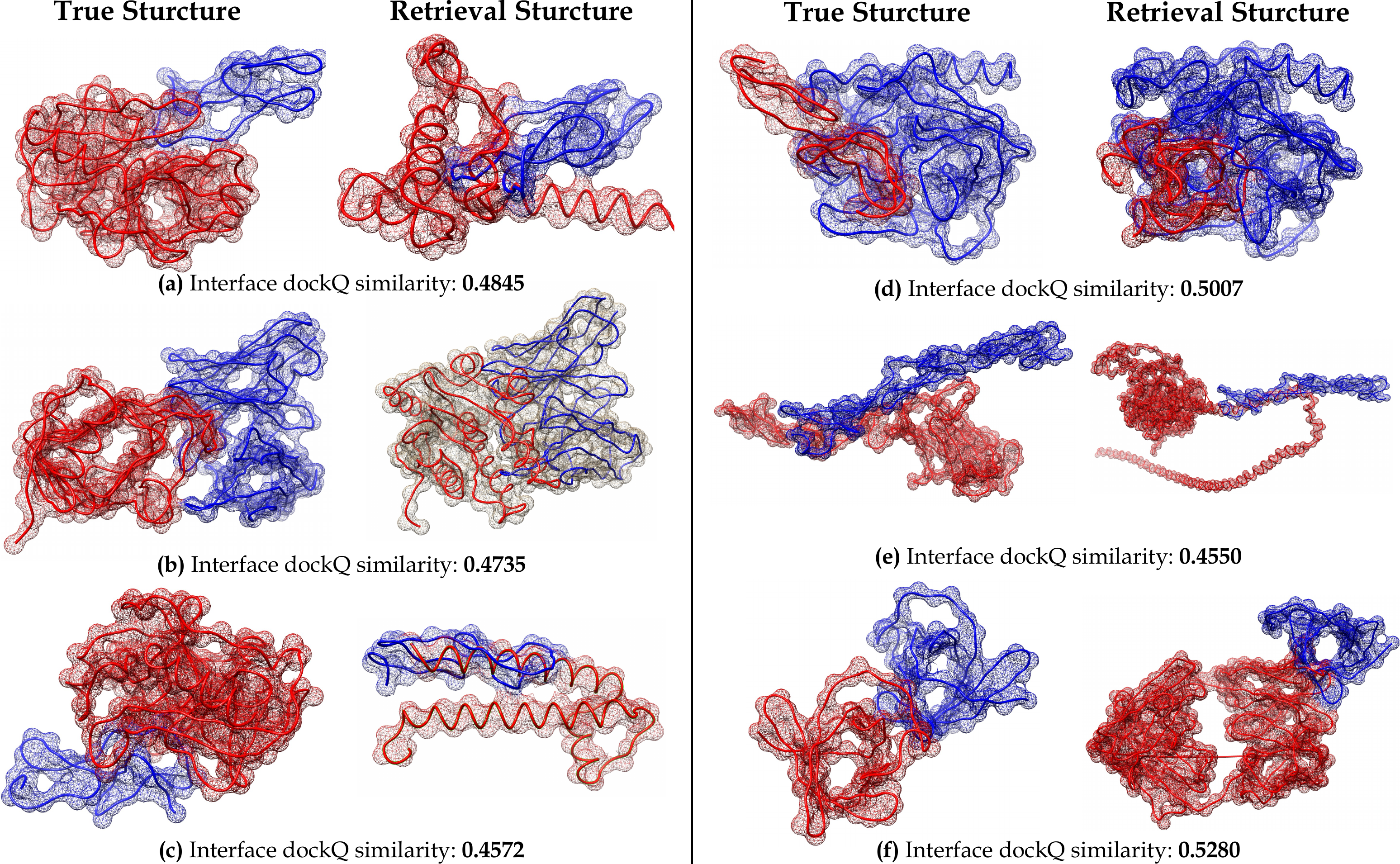}}
\vspace{-0.5cm}
  \caption{Visualization of PPIretrieval results for proteins in the PDB test set, evaluated by \textit{dockQ}. Proteins colored in blue are input query proteins; proteins colored in red are binding partners. Left column displays the ground-truth structures; right column shows the structures predicted by PPIretrieval.
  \vspace{-0.4cm}
  \label{fig:search.visualization}
}
\end{figure}

In Fig.~\ref{fig:search.visualization}, we observe that the predicted PPIs, with interface \textit{dockQ} similarity, form a well-defined \textit{lock-and-key} structure. This reliable structure formation bolsters confidence in the potential of PPIretrieval for exploring novel protein interactions. One can utilize our model and database to investigate and learn about unknown protein interactions. Additional visualizations are available in App.~\ref{appen:surface.visualization}.

\vspace{-0.3cm}
\section{Experiment}
\vspace{-0.2cm}
\label{sec:experiment}

We use the PDB dataset from \citep{gainza2020deciphering, sverrisson2021fast} and the DIPS dataset from \citep{morehead2023dips}. Also, we clean up PPI data from the PPBS dataset from \citep{tubiana2022scannet}. To ensure data quality, we exclude protein complexes with fewer than $40$ residues in one protein or a minimum distance between two proteins exceeding $10\mathring{\text{A}}$. The original PDB dataset comprises $4754$ and $933$ protein complexes for training and testing, respectively. After excluding low-quality complexes, the PDB dataset consists of $4420$ and $840$ protein complexes for training and testing, with 10\% of the training set used for validation following \citep{sverrisson2021fast}. The original DIPS dataset includes $33159$ and $8290$ protein complexes for training and validation, respectively. After excluding low-quality complexes, the DIPS dataset consists of $33028$ and $8267$ protein complexes for training and validation, with the first $4267$ complexes of the validation set used for testing. The original PPBS dataset comprises $101755$, $10221$, and $10911$ protein complexes for training, validation, and testing, respectively (following homology split in \citep{tubiana2022scannet}). After excluding low-quality complexes, the PPBS dataset consists of $99799$, $9960$, and $10742$ protein complexes for training, validation, and testing.

For surface sampling of each protein, we use dMaSIF \citep{sverrisson2021fast} with sampling resolution $1.0\mathring{\text{A}}$, sampling distance $2.25\mathring{\text{A}}$, sampling number $20$, and sampling variance $0.3\mathring{\text{A}}$. Additionally, we use $32$ hidden dims and $0.3$ dropout for all projections, use $2$ propagation layers and $2$ cross-attention layers. We choose a learning rate of $1e-4$ and use the AdamW optimizer with a weight decay of $5e-10$.
We select the models with the lowest validation loss $\mathcal{L}$.

\begin{figure}[ht!]
\vspace{-0.3cm}
\centering
{
\includegraphics[width=.8\columnwidth]{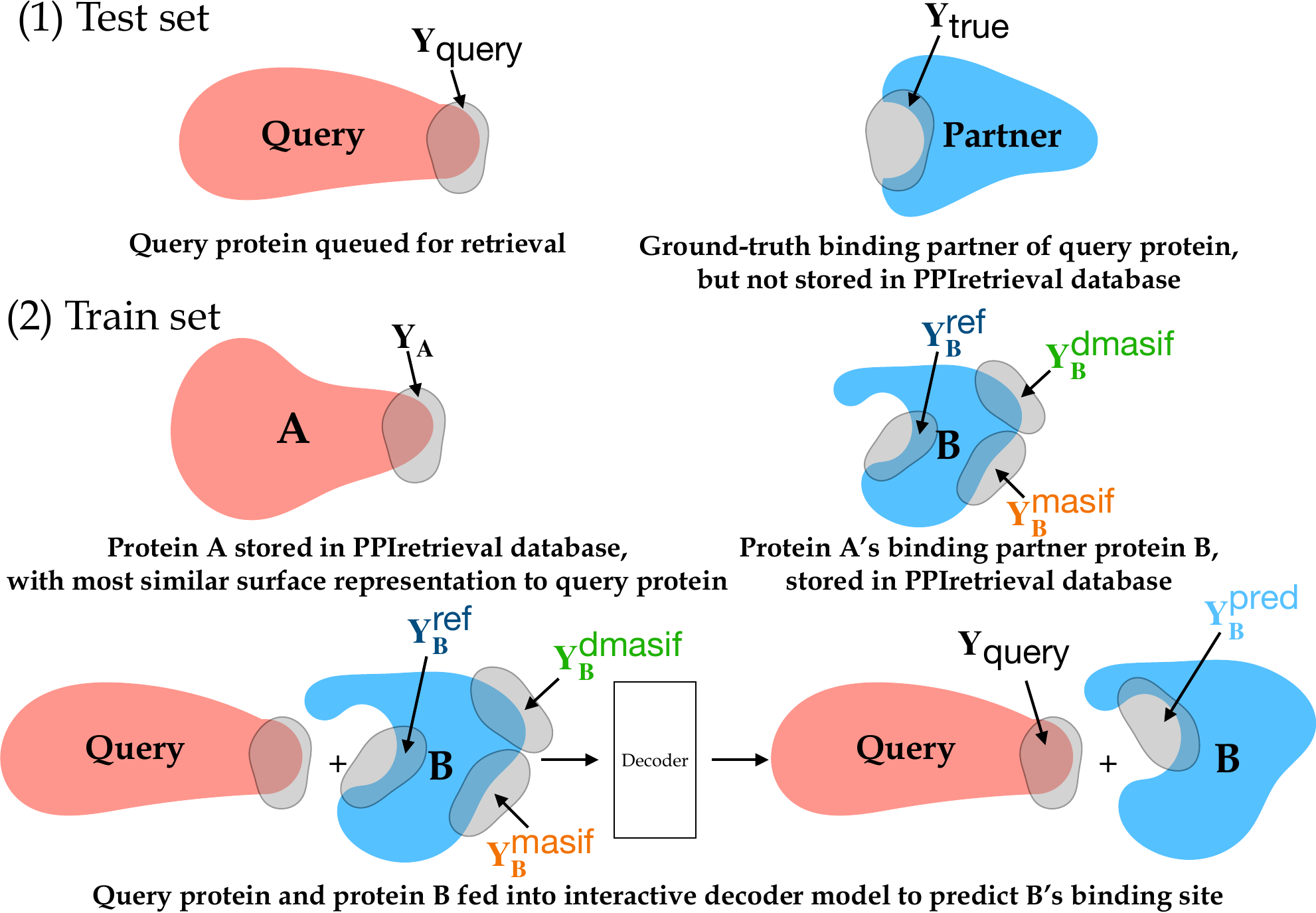}}
\vspace{-0.3cm}
  \caption{Evaluation of PPI binding site during inference. For a PPI in the test set, a query protein with a known binding site $\mathbf{Y}_\text{query}$ seeks a binding partner with an actual binding site $\mathbf{Y}_\text{true}$. 
  However, we assume that the binding partner is unknown to us.
  So, PPIretrieval aims to retrieve a potential binding partner from the surface databse.
  PPIretrieval identifies protein $A$ in the surface database, which has the most similar surface representation to the query protein. Protein $A$ has a known binding partner $B$ with a reference binding site $\mathbf{Y}_B^{\text{ref}}$ (stored in database), a binding site $\mathbf{Y}_B^{\text{masif}}$ predicted by masif, and a binding site $\mathbf{Y}_B^{\text{dmasif}}$ predicted by dmasif. 
  PPIretrieval takes query protein and $B$ as input and predicts a new binding site $\mathbf{Y}_B^{\text{pred}}$.  We compute 
  $\textit{dockQ}(\mathbf{Y}_\text{true}, \mathbf{Y}_B^{\text{pred}}), \textit{TM}(\mathbf{Y}_\text{true}, \mathbf{Y}_B^{\text{pred}}), \textit{rmsd}(\mathbf{Y}_\text{true}, \mathbf{Y}_B^{\text{pred}})$, 
  $\textit{dockQ}(\mathbf{Y}_\text{true}, \mathbf{Y}_B^{\text{masif}}), \textit{TM}(\mathbf{Y}_\text{true}, \mathbf{Y}_B^{\text{masif}}), \textit{rmsd}(\mathbf{Y}_\text{true}, \mathbf{Y}_B^{\text{masif}})$, 
  $\textit{dockQ}(\mathbf{Y}_\text{true}, \mathbf{Y}_B^{\text{dmasif}}), \textit{TM}(\mathbf{Y}_\text{true}, \mathbf{Y}_B^{\text{dmasif}}), \textit{rmsd}(\mathbf{Y}_\text{true}, \mathbf{Y}_B^{\text{dmasif}})$,
  $\textit{dockQ}(\mathbf{Y}_\text{true}, \mathbf{Y}_B^{\text{ref}}), \textit{TM}(\mathbf{Y}_\text{true}, \mathbf{Y}_B^{\text{ref}}), \textit{rmsd}(\mathbf{Y}_\text{true}, \mathbf{Y}_B^{\text{ref}})$ to evaluate and compare the quality of PPI and binding interfaces. $\mathbf{Y}_\text{true}$ denotes the known binding site of the ground-truth binding partner; $\mathbf{Y}^\text{ref}_B$ denotes the known binding site (stored in database) of the retrieved binding partner; $\mathbf{Y}^\text{pred}_B$ denotes the predicted binding site of the retrieved binding partner.
  \label{fig:emperical.eval}
  \vspace{-0.7cm}
}
\end{figure}

\vspace{-0.2cm}
\subsection{Empirical Evaluation}
\vspace{-0.2cm}
We empirically assess the quality of PPIs carried out by PPIretrieval during the inference stage. For PPIs in the test set, we measure the \textit{dockQ} score \citep{basu2016dockq}, \textit{TM} score \citep{zhang2005tm}, and root-mean-square-distance (\textit{rmsd}) to compare PPIretrieval reference binding site, masif-predicted binding site \citep{gainza2020deciphering}, dmasif-predicted binding site \citep{sverrisson2021fast}, and PPIretrieval predicted binding site with the ground-truth binding sites. (1) \textbf{\textit{dockQ} score} measures the quality between a ground-truth binding site and a predicted binding site, which combines the fraction of native contacts, the interface root mean square distance, and the ligand root mean square distance; a higher \textit{dockQ} score indicates a better quality of the predicted binding site. (2) \textbf{\textit{TM} score} measures the similarity between a ground-truth binding site and a predicted binding site, which considers both the distance and the alignment of the residues; a higher \textit{TM} score indicates higher similarity between the two binding site. (3) \textbf{\textit{rmsd}} measures the distance between a ground-truth binding site and a predicted binding site after superimposition; a lower \textit{rmsd} indicates higher superimposition similarity between the two binding site. A comprehensive visualization and explanation of our metrics to assess quality of PPI binding sites are demonstrated in Fig.~\ref{fig:emperical.eval}.

\noindent \textbf{Inference Result} \quad
In Tab.~\ref{tab:ppisearch.baseline1}, we assess the quality of PPIs and binding interface identified by PPIretrieval with smaller databases. The database for each test set comprises surface representations from the training set of the respective dataset. For example, when evaluating the PDB test set, we only search for surface representations in the PDB training set. Also, the models are trained on each respective dataset.
\begin{table}[htbp!]
\footnotesize
  \centering
  \vspace{-0.3cm}
  \resizebox{.8\columnwidth}{!}{%
\begin{tabular}{c|cc|c|c|c}
\hline
\hline
\rowcolor[rgb]{ .851,  .851,  .851} \textbf{Dataset} & \multicolumn{2}{c|}{\textbf{Metrics}} & \textbf{PDB} & \textbf{DIPS} & \textbf{PPBS} \\ 
\hline
\multirow{12}{*}{Site Quality} & \multicolumn{1}{c|}{\multirow{4}{*}{\textit{dockQ}$(\uparrow)$}} & \cellcolor[rgb]{ .851,  .882, .949} $\mathbf{Y}_\text{true}, \mathbf{Y}^\text{ref}_B$ &  $0.4073$   &  $0.4177$    &   $0.5535$    \\
& \multicolumn{1}{c|}{}& \cellcolor[rgb]{ .706,  .776,  .906} $\mathbf{Y}_\text{true}, \mathbf{Y}^\text{pred}_B$  &  $\underline{0.4220}$   & $\underline{0.4304}$     &   $\underline{0.5946}$   \\ 
& \multicolumn{1}{c|}{}& \cellcolor[rgb]{ .6,  .6,  .8} $\mathbf{Y}_\text{true}, \mathbf{Y}^\text{masif}_B$ &  $0.1334$   &  $0.1021$    &   $0.1228$   \\ 
& \multicolumn{1}{c|}{}& \cellcolor[rgb]{ .45,  .45,  .7} $\mathbf{Y}_\text{true}, \mathbf{Y}^\text{dmasif}_B$ &  $0.1155$   &  $0.0837$    &   $0.1036$  \\
\cline{2-6} 
& \multicolumn{1}{c|}{\multirow{4}{*}{\textit{TM}$(\uparrow)$}}& \cellcolor[rgb]{ .851,  .882, .949} $\mathbf{Y}_\text{true}, \mathbf{Y}^\text{ref}_B$ &  ${0.2134}$   &  $0.6617$     &  ${0.4622}$   \\
& \multicolumn{1}{c|}{}& \cellcolor[rgb]{ .706,  .776,  .906} $\mathbf{Y}_\text{true}, \mathbf{Y}^\text{pred}_B$ &  $\underline{0.2366}$   &  $\underline{0.6649}$    &   $\underline{0.4735}$     \\
& \multicolumn{1}{c|}{}& \cellcolor[rgb]{ .6,  .6,  .8} $\mathbf{Y}_\text{true}, \mathbf{Y}^\text{masif}_B$ &  $0.0773$   &  $0.0981$    &   $0.0911$   \\
& \multicolumn{1}{c|}{}& \cellcolor[rgb]{ .45,  .45,  .7} $\mathbf{Y}_\text{true}, \mathbf{Y}^\text{dmasif}_B$ &  $0.0665$   &  $0.0831$    &   $0.0871$  \\
\cline{2-6} 
& \multicolumn{1}{c|}{\multirow{4}{*}{\textit{rmsd}$(\downarrow)$}}& \cellcolor[rgb]{ .851,  .882, .949} $\mathbf{Y}_\text{true}, \mathbf{Y}^\text{ref}_B$ &  ${11.40}$   &  $11.33$     &  $\underline{8.20}$   \\
& \multicolumn{1}{c|}{}& \cellcolor[rgb]{ .706,  .776,  .906} $\mathbf{Y}_\text{true}, \mathbf{Y}^\text{pred}_B$ &  $\underline{10.44}$   &  $\underline{6.02}$   &   $9.77$ \\ 
& \multicolumn{1}{c|}{}& \cellcolor[rgb]{ .6,  .6,  .8} $\mathbf{Y}_\text{true}, \mathbf{Y}^\text{masif}_B$ &  $15.73$   &  $19.66$    &   $17.32$   \\ 
& \multicolumn{1}{c|}{}& \cellcolor[rgb]{ .45,  .45,  .7} $\mathbf{Y}_\text{true}, \mathbf{Y}^\text{dmasif}_B$ &  $17.87$   &  $23.55$    &   $19.65$  \\
\hline
\hline
\end{tabular}
}
\vspace{-0.3cm}
    \caption{\textit{dockQ}, \textit{TM}, and \textit{rmsd} for evaluation of \textbf{Top1 hit} binding sites predicted by PPIretrieval in comparison with other binding sites over three runs. The database for each test set comprises surface representations from the training and validation sets of each respective dataset.}
    \vspace{-0.5cm}
  \label{tab:ppisearch.baseline1}%
\end{table}

In Tab.~\ref{tab:ppisearch.baseline2}, we evaluate the quality of PPIs and binding interface identified by PPIretrieval with larger database. The database includes surface representations from the training and validation sets of the PDB, DIPS, and PPBS datasets, in total of $155,384$ paired proteins with their surface features for retrieval. And the models are trained on all training PPIs.
\begin{table}[htbp!]
\footnotesize
  \centering
  \vspace{-0.5cm}
  \resizebox{.8\columnwidth}{!}{%
\begin{tabular}{c|cc|c|c|c}
\hline
\hline
\rowcolor[rgb]{ .851,  .851,  .851} \textbf{Dataset} & \multicolumn{2}{c|}{\textbf{Metrics}} & \textbf{PDB} & \textbf{DIPS} & \textbf{PPBS} \\ \hline
\multirow{12}{*}{Site Quality} & \multicolumn{1}{c|}{\multirow{4}{*}{\textit{dockQ}$(\uparrow)$}} & \cellcolor[rgb]{ .851,  .882, .949} $\mathbf{Y}_\text{true}, \mathbf{Y}^\text{ref}_B$ & $0.4250$    &  ${0.4367}$    &  $0.6014$    \\
& \multicolumn{1}{c|}{}& \cellcolor[rgb]{ .706,  .776,  .906} $\mathbf{Y}_\text{true}, \mathbf{Y}^\text{pred}_B$  & $\underline{0.4678}$    &  $\underline{0.4410}$    &  $\underline{0.6045}$    \\ 
& \multicolumn{1}{c|}{}& \cellcolor[rgb]{ .6,  .6,  .8} $\mathbf{Y}_\text{true}, \mathbf{Y}^\text{masif}_B$ &  $0.1345$   &  $0.1026$    &   $0.1249$   \\ 
& \multicolumn{1}{c|}{}& \cellcolor[rgb]{ .45,  .45,  .7} $\mathbf{Y}_\text{true}, \mathbf{Y}^\text{dmasif}_B$ &  $0.1235$   &  $0.0998$    &   $0.1261$  \\
\cline{2-6} 
& \multicolumn{1}{c|}{\multirow{4}{*}{\textit{TM}$(\uparrow)$}} & \cellcolor[rgb]{ .851,  .882, .949}$\mathbf{Y}_\text{true}, \mathbf{Y}^\text{ref}_B$ & $0.3222$    &  ${0.6914}$    &  $0.6014$    \\
& \multicolumn{1}{c|}{}& \cellcolor[rgb]{ .706,  .776,  .906} $\mathbf{Y}_\text{true}, \mathbf{Y}^\text{pred}_B$  & $\underline{0.3300}$    &  $\underline{0.6944}$    &  $\underline{0.6045}$    \\
& \multicolumn{1}{c|}{}& \cellcolor[rgb]{ .6,  .6,  .8} $\mathbf{Y}_\text{true}, \mathbf{Y}^\text{masif}_B$ &  $0.0833$   &  $0.1002$    &   $0.1004$   \\ 
& \multicolumn{1}{c|}{}& \cellcolor[rgb]{ .45,  .45,  .7} $\mathbf{Y}_\text{true}, \mathbf{Y}^\text{dmasif}_B$ &  $0.0823$   &  $0.0911$    &   $0.1144$  \\
\cline{2-6} 
& \multicolumn{1}{c|}{\multirow{3}{*}{\textit{rmsd}$(\downarrow)$}}& \cellcolor[rgb]{ .851,  .882, .949}$\mathbf{Y}_\text{true}, \mathbf{Y}^\text{ref}_B$ & $\underline{9.30}$    &   $9.65$   &  $9.96$     \\
& \multicolumn{1}{c|}{}& \cellcolor[rgb]{ .706,  .776,  .906}$\mathbf{Y}_\text{true}, \mathbf{Y}^\text{pred}_B$ &  $10.70$   &  $\underline{5.67}$    &   $\underline{6.52}$  \\ 
& \multicolumn{1}{c|}{}& \cellcolor[rgb]{ .6,  .6,  .8} $\mathbf{Y}_\text{true}, \mathbf{Y}^\text{masif}_B$ &  $15.56$   &  $19.05$    &   $16.82$   \\ 
& \multicolumn{1}{c|}{}& \cellcolor[rgb]{ .45,  .45,  .7} $\mathbf{Y}_\text{true}, \mathbf{Y}^\text{dmasif}_B$ &  $16.81$   &  $21.22$    &   $16.08$  \\
\hline
\hline
\end{tabular}
}
\vspace{-0.3cm}
    \caption{\textit{dockQ}, \textit{TM}, and \textit{rmsd} for evaluation of \textbf{Top1 hit} binding sites predicted by PPIretrieval in comparison with other binding sites over three runs. The database comprises all surface representations from the training and validation sets of PDB, DIPS, and PPBS datasets.}
    \vspace{-0.5cm}
  \label{tab:ppisearch.baseline2}%
\end{table}

By comparing the tabular results, it is evident that the qualities of the predicted PPIs carried out by PPIretrieval improve with a larger database, as reflected in higher \textit{dockQ} and \textit{TM} scores, as well as lower \textit{rmsd}. This improvement suggests that using PPIretrieval could be highly beneficial in facilitating the discovery of novel PPIs. More experimental results can be found in App.~\ref{appen:baseline.models}

\noindent \textbf{Cross-Dataset Validation} \quad
In addition, we show the cross-dataset performance in Tab.~\ref{tab:experiment.cross.dataset}. We take the model trained on PDB training set only to encode the PPIs in DIPS and PPBS training and validation sets, respectively. Then we evaluate the cross-dataset performance on DIPS and PPBS test sets at two different hit rates, respectively. \textbf{Top10 hit} means that PPIretrieval retrieves the $10$ most similar surface representations to the query protein in the test set for inference. Then, PPIretrieval decodes between the query protein and the $10$ potential binding partners associated with these similar proteins. The best binding partner for the query protein is then selected based on the highest \textit{dockQ} score.
\begin{table}[htbp!]
\footnotesize
  \centering
  \vspace{-0.3cm}
  \resizebox{\columnwidth}{!}{%
\begin{tabular}{c|cc|c|c|c|c}
\hline
\hline
\rowcolor[rgb]{ .851,  .851,  .851} \textbf{Dataset} & \multicolumn{2}{c|}{\textbf{Metrics}} & \textbf{DIPS-Top1} & \textbf{DIPS-Top10} & \textbf{PPBS-Top1} & \textbf{PPBS-Top10} \\ \hline
\multirow{12}{*}{Site Quality} & \multicolumn{1}{c|}{\multirow{4}{*}
{\textit{dockQ}$(\uparrow)$}} & \cellcolor[rgb]{ .851,  .882, .949} 
$\mathbf{Y}_\text{true}, \mathbf{Y}^\text{ref}_B$ & $0.4030$   &  $0.4156$    & $0.5231$ &  $0.5611$  \\
& \multicolumn{1}{c|}{}& \cellcolor[rgb]{ .706,  .776,  .906} $\mathbf{Y}_\text{true}, \mathbf{Y}^\text{pred}_B$  &  $\underline{0.4207}$   &  $\underline{0.4435}$    & $\underline{0.5579}$  & $\underline{0.5857}$   \\   & \multicolumn{1}{c|}{}& \cellcolor[rgb]{ .6,  .6,  .8} $\mathbf{Y}_\text{true}, \mathbf{Y}^\text{masif}_B$ &  $0.0515$   &  $0.0523$    &   $0.0621$  &   $0.0633$ \\ 
& \multicolumn{1}{c|}{}& \cellcolor[rgb]{ .45,  .45,  .7} $\mathbf{Y}_\text{true}, \mathbf{Y}^\text{dmasif}_B$ &  $0.0434$   &  $0.0515$    &   $0.0601$  &   $0.0633$  \\
\cline{2-7} 
& \multicolumn{1}{c|}{\multirow{4}{*}
{\textit{TM}$(\uparrow)$}} & \cellcolor[rgb]{ .851,  .882, .949} $\mathbf{Y}_\text{true}, \mathbf{Y}^\text{ref}_B$ & $0.5330$   &  $0.6714$    & $0.4202$ &  $0.3725$  \\
& \multicolumn{1}{c|}{}& \cellcolor[rgb]{ .706,  .776,  .906} $\mathbf{Y}_\text{true}, \mathbf{Y}^\text{pred}_B$  &  $\underline{0.5419}$   &  $\underline{0.6792}$    & $\underline{0.4421}$  & $\underline{0.3889}$   \\  & \multicolumn{1}{c|}{}& \cellcolor[rgb]{ .6,  .6,  .8} $\mathbf{Y}_\text{true}, \mathbf{Y}^\text{masif}_B$ &  $0.0499$   &  $0.0511$    &   $0.0611$  &   $0.0620$ \\ 
& \multicolumn{1}{c|}{}& \cellcolor[rgb]{ .45,  .45,  .7} $\mathbf{Y}_\text{true}, \mathbf{Y}^\text{dmasif}_B$ &  $0.0433$   &  $0.0491$    &   $0.0519$  &   $0.0602$  \\
\cline{2-7} 
& \multicolumn{1}{c|}{\multirow{4}{*}{\textit{rmsd}$(\downarrow)$}}& \cellcolor[rgb]{ .851,  .882, .949} $\mathbf{Y}_\text{true}, \mathbf{Y}^\text{ref}_B$ &  $11.12$   &  $\underline{7.35}$    & $\underline{8.92}$  & $\underline{8.91}$ \\
& \multicolumn{1}{c|}{}& \cellcolor[rgb]{ .706,  .776,  .906} $\mathbf{Y}_\text{true}, \mathbf{Y}^\text{pred}_B$ &  $\underline{5.84}$   &  $10.50$    &   $11.76$ & $10.49$ \\ 
& \multicolumn{1}{c|}{}& \cellcolor[rgb]{ .6,  .6,  .8} $\mathbf{Y}_\text{true}, \mathbf{Y}^\text{masif}_B$ &  $20.73$   &  $19.21$    &   $19.81$  &   $19.55$    \\ 
& \multicolumn{1}{c|}{}& \cellcolor[rgb]{ .45,  .45,  .7} $\mathbf{Y}_\text{true}, \mathbf{Y}^\text{dmasif}_B$ &  $23.89$   &  $22.05$    &   $20.66.$  &   $20.04$  \\
\hline
\hline
\end{tabular}
}
\vspace{-0.4cm}
    \caption{\textit{dockQ}, \textit{TM}, and \textit{rmsd} for evaluation of \textbf{Top1, Top10 hit} binding sites predicted by PPIretrieval in comparison with other binding sites on cross-datasets over three runs. The database for each test set comprises surface representations from the training and validation sets of each respective dataset.}
    \vspace{-0.5cm}
  \label{tab:experiment.cross.dataset}%
\end{table}

We observe solid cross-dataset validation results, with improvements in \textit{dockQ} scores compared to the baseline results in Table~\ref{tab:ppisearch.baseline1}. This improvement indicates that our model and training strategy demonstrate generalizability to unseen PPI structures. Furthermore, the trained models can be directly employed to encode new PPIs and conduct inference without the need for retraining. We provide more comparison in Tab.~\ref{tab:pdb.vs.ppbs.crossdataset} and visualization in Fig.~\ref{fig:pdb.ppbs.crossdataset}.

\noindent \textbf{Computational Resources} \quad
Our models are trained on a single Nvidia 48G A40 GPU. Regarding training time, PPIretrieval takes approximately $0.35$s to train a protein complex. In terms of inference time, PPIretrieval requires about $0.11$s for a protein complex.

\vspace{-0.2cm}
\subsection{Ablation Study}
\vspace{-0.2cm}
In Tab.~\ref{tab:topk.hit.ablation}, we present experimental results for PPIretrieval at five different hit rates, increasing from \textbf{Top1} to \textbf{Top100}. The models are trained on PDB training set only.
\begin{table}[htbp!]
\footnotesize
  \centering
  \vspace{-0.3cm}
  \resizebox{\columnwidth}{!}{%
\begin{tabular}{c|cc|c|c|c|c|c}
\hline
\hline
\rowcolor[rgb]{ .851,  .851,  .851} \textbf{PDB Dataset} & \multicolumn{2}{c|}{\textbf{Metrics}} & \textbf{Top1} & \textbf{Top10} & \textbf{Top20} & \textbf{Top50} & \textbf{Top100} \\ \hline
\multirow{12}{*}{Site Quality} & \multicolumn{1}{c|}{\multirow{4}{*}{\textit{dockQ}$(\uparrow)$}} & \cellcolor[rgb]{ .851,  .882, .949} $\mathbf{Y}_\text{true}, \mathbf{Y}^\text{ref}_B$ &  $0.4073$   &  $0.4362$  & $0.4379$  & $0.4411$ & $0.3507$  \\
& \multicolumn{1}{c|}{}& \cellcolor[rgb]{ .706,  .776,  .906} $\mathbf{Y}_\text{true}, \mathbf{Y}^\text{pred}_B$  &  $\underline{0.4220}$   &  $\underline{0.4375}$ & $\underline{0.4459}$ & $\underline{0.4569}$ &  $\cellcolor[rgb]{ .573,  .816,  .314}\underline{0.4688}$ \\ 
& \multicolumn{1}{c|}{}& \cellcolor[rgb]{ .6,  .6,  .8} $\mathbf{Y}_\text{true}, \mathbf{Y}^\text{masif}_B$ &  $0.1334$   &  $0.1338$    &   $0.1355$  &   $0.1401$  &   $0.1405$ \\ 
& \multicolumn{1}{c|}{}& \cellcolor[rgb]{ .45,  .45,  .7} $\mathbf{Y}_\text{true}, \mathbf{Y}^\text{dmasif}_B$ &  $0.1155$   &  $0.1194$    &   $0.1212$  &   $0.1247$ &   $0.1301$ \\
\cline{2-8} 
& \multicolumn{1}{c|}{\multirow{4}{*}{\textit{TM}$(\uparrow)$}}& \cellcolor[rgb]{ .851,  .882, .949} $\mathbf{Y}_\text{true}, \mathbf{Y}^\text{ref}_B$ &  $0.2134$   &  $0.2078$    & $0.2059$  & $0.2059$ & $0.2108$  \\
& \multicolumn{1}{c|}{}& \cellcolor[rgb]{ .706,  .776,  .906} $\mathbf{Y}_\text{true}, \mathbf{Y}^\text{pred}_B$ &  $\underline{0.2366}$   &  $\underline{0.2266}$    & $\underline{0.2241}$ & $\underline{0.2231}$ &  $\underline{0.2265}$   \\ 
& \multicolumn{1}{c|}{}& \cellcolor[rgb]{ .6,  .6,  .8} $\mathbf{Y}_\text{true}, \mathbf{Y}^\text{masif}_B$ &  $0.0773$   &  $0.0775$    &   $0.0774$  &   $0.0758$  &   $0.0702$ \\ 
& \multicolumn{1}{c|}{}& \cellcolor[rgb]{ .45,  .45,  .7} $\mathbf{Y}_\text{true}, \mathbf{Y}^\text{dmasif}_B$ &  $0.0665$   &  $0.0668$    &   $0.0679$  &   $0.0698$ &   $0.0701$ \\
\cline{2-8} 
& \multicolumn{1}{c|}{\multirow{4}{*}{\textit{rmsd}$(\downarrow)$}}& \cellcolor[rgb]{ .851,  .882, .949} $\mathbf{Y}_\text{true}, \mathbf{Y}^\text{ref}_B$ &  ${11.40}$   &  $9.74$    & $9.59$  & ${9.50}$ & $9.34$  \\
& \multicolumn{1}{c|}{}& \cellcolor[rgb]{ .706,  .776,  .906} $\mathbf{Y}_\text{true}, \mathbf{Y}^\text{pred}_B$ &  $\underline{10.44}$   &  $\underline{9.33}$    & $\underline{8.94}$ & $\underline{8.52}$ &  $ \cellcolor[rgb]{ .573,  .816,  .314}\underline{8.16}$  \\ 
& \multicolumn{1}{c|}{}& \cellcolor[rgb]{ .6,  .6,  .8} $\mathbf{Y}_\text{true}, \mathbf{Y}^\text{masif}_B$ &  $15.98$   &  $15.88$    &   $15.84$  &   $15.76$ &   $15.53$ \\
& \multicolumn{1}{c|}{}& \cellcolor[rgb]{ .45,  .45,  .7} $\mathbf{Y}_\text{true}, \mathbf{Y}^\text{dmasif}_B$ &  $17.87$   &  $17.31$    &   $17.03$  &   $16.55$ &   $16.02$ \\
\hline
Cost & \multicolumn{1}{c|}{\textit{PPIretrieval runtime}($\downarrow$)}& second/protein &  $0.29$   &  $0.91$    & $1.97$  & $4.64$ & $9.44$  \\ 
\hline
\hline
\end{tabular}
}
\vspace{-0.4cm}
    \caption{\textit{dockQ}, \textit{TM}, and \textit{rmsd} for evaluation of \textbf{Top1, Top10, Top20, Top50, Top100 hit} binding sites predicted by PPIretrieval in comparison with other binding sites in the PDB test set over three runs. The database comprises surface features from training and validation sets from PDB dataset only.}
    \vspace{-0.5cm}
  \label{tab:topk.hit.ablation}%
\end{table}

We observe an improvement in the quality of predicted PPIs, measured by \textit{dockQ, TM} scores and \textit{rmsd}, as the hit rate increases from \textbf{Top1} to \textbf{Top100}. Notably, with 100 similar representations, the predicted PPIs exhibit high quality in terms of the \textit{dockQ} and \textit{TM} scores. This suggests potential PPI exploration with PPIretrieval.

Furthermore, we present experimental results for PPIretrieval at five different hit rates, increasing from \textbf{Top1} to \textbf{Top100}, in Tab.~\ref{tab:topk.hit.ablation2}. The models are trained on all PDB, DIPS, and PPBS training set, and the database comprises surface representations from training and validation sets of them.
\begin{table}[htbp!]
\footnotesize
  \centering
  \vspace{-0.3cm}
  \resizebox{\columnwidth}{!}{%
\begin{tabular}{c|cc|c|c|c|c|c}
\hline
\hline
\rowcolor[rgb]{ .851,  .851,  .851} \textbf{PDB Dataset} & \multicolumn{2}{c|}{\textbf{Metrics}} & \textbf{Top1} & \textbf{Top10} & \textbf{Top20} & \textbf{Top50} & \textbf{Top100} \\ \hline
\multirow{12}{*}{Site Quality} & \multicolumn{1}{c|}{\multirow{4}{*}{\textit{dockQ}$(\uparrow)$}} & \cellcolor[rgb]{ .851,  .882, .949} $\mathbf{Y}_\text{true}, \mathbf{Y}^\text{ref}_B$ &  $0.4126$   &  $0.4331$  & $0.4480$  & $0.4491$ & $0.4490$  \\
& \multicolumn{1}{c|}{}& \cellcolor[rgb]{ .706,  .776,  .906} $\mathbf{Y}_\text{true}, \mathbf{Y}^\text{pred}_B$  &  $\underline{0.4235}$   &  $\underline{0.4402}$ & $\underline{0.4531}$ & $\underline{0.4649}$ &  $\cellcolor[rgb]{ .573,  .816,  .314}\underline{0.4708}$  \\ 
& \multicolumn{1}{c|}{}& \cellcolor[rgb]{ .6,  .6,  .8} $\mathbf{Y}_\text{true}, \mathbf{Y}^\text{masif}_B$ &  $0.1433$   &  $0.1436$    &   $0.1455$  &   $0.1458$ &   $0.1478$ \\ 
& \multicolumn{1}{c|}{}& \cellcolor[rgb]{ .45,  .45,  .7} $\mathbf{Y}_\text{true}, \mathbf{Y}^\text{dmasif}_B$ &  $0.1225$   &  $0.1266$    &   $0.1301$  &   $0.1398$ &   $0.1405$ \\
\cline{2-8} 
& \multicolumn{1}{c|}{\multirow{4}{*}{\textit{TM}$(\uparrow)$}}& \cellcolor[rgb]{ .851,  .882, .949} $\mathbf{Y}_\text{true}, \mathbf{Y}^\text{ref}_B$ &  $0.3944$   &  $0.3877$    & $0.3833$  & $0.3554$ & $0.3422$  \\
& \multicolumn{1}{c|}{}& \cellcolor[rgb]{ .706,  .776,  .906} $\mathbf{Y}_\text{true}, \mathbf{Y}^\text{pred}_B$ &  $\underline{0.4041}$   &  $\underline{0.3969}$    & $\underline{0.3863}$ & $\underline{0.3625}$ &  $\underline{0.3528}$  \\ 
& \multicolumn{1}{c|}{}& \cellcolor[rgb]{ .6,  .6,  .8} $\mathbf{Y}_\text{true}, \mathbf{Y}^\text{masif}_B$ &  $0.0787$   &  $0.0766$    &   $0.0750$  &   $0.0721$ &   $0.0709$  \\ 
& \multicolumn{1}{c|}{}& \cellcolor[rgb]{ .45,  .45,  .7} $\mathbf{Y}_\text{true}, \mathbf{Y}^\text{dmasif}_B$ &  $0.0536$   &  $0.0548$    &   $0.0588$  &   $0.0582$ &   $0.0601$ \\
\cline{2-8} 
& \multicolumn{1}{c|}{\multirow{4}{*}{\textit{rmsd}$(\downarrow)$}}& \cellcolor[rgb]{ .851,  .882, .949} $\mathbf{Y}_\text{true}, \mathbf{Y}^\text{ref}_B$ &  ${10.41}$   &  $9.73$    & $9.70$  & $9.49$ & $9.32$  \\
& \multicolumn{1}{c|}{}& \cellcolor[rgb]{ .706,  .776,  .906} $\mathbf{Y}_\text{true}, \mathbf{Y}^\text{pred}_B$ &  $\underline{10.04}$   &  $\underline{8.97}$    & $\underline{8.66}$ & $\underline{8.20}$ &  $ \cellcolor[rgb]{ .573,  .816,  .314}\underline{7.35}$  \\ 
& \multicolumn{1}{c|}{}& \cellcolor[rgb]{ .6,  .6,  .8} $\mathbf{Y}_\text{true}, \mathbf{Y}^\text{masif}_B$ &  $15.73$   &  $15.71$    &   $15.54$  &   $15.26$  &   $15.19$ \\ 
& \multicolumn{1}{c|}{}& \cellcolor[rgb]{ .45,  .45,  .7} $\mathbf{Y}_\text{true}, \mathbf{Y}^\text{dmasif}_B$ &  $17.75$   &  $17.22$    &   $17.02$  &   $16.40$ &   $16.11$ \\
\hline
\hline
\end{tabular}
}
\vspace{-0.4cm}
    \caption{\textit{dockQ}, \textit{TM}, and \textit{rmsd} for evaluation of \textbf{Top1, Top10, Top20, Top50, Top100 hit} binding sites predicted by PPIretrieval in comparison with other binding sites in the PDB test set over three runs. The database comprises surface features from training and validation sets from PDB, DIPS, and PPBS dataset.}
    \vspace{-0.5cm}
  \label{tab:topk.hit.ablation2}%
\end{table}

We observe improved predicted interface quality in terms of \textit{dockQ, TM} scores,  and \textit{rmsd}, with larger surface database. The robust experimental results suggest that PPIretrieval has the potential to facilitate and expedite the discovery of novel PPIs, identifying candidates with higher \textit{dockQ} scores. However, it is important to note that the computational time has also increased significantly. As a retrieval model, there exists a trade-off between performance and efficiency, and this trade-off becomes evident with higher hit rates. More ablation studies can be found in App.~\ref{appen:baseline.models}

\vspace{-0.2cm}
\subsection{Case Study}
\begin{figure}[ht!]
\vspace{-0.3cm}
\centering
{
\includegraphics[width=1.05\columnwidth]{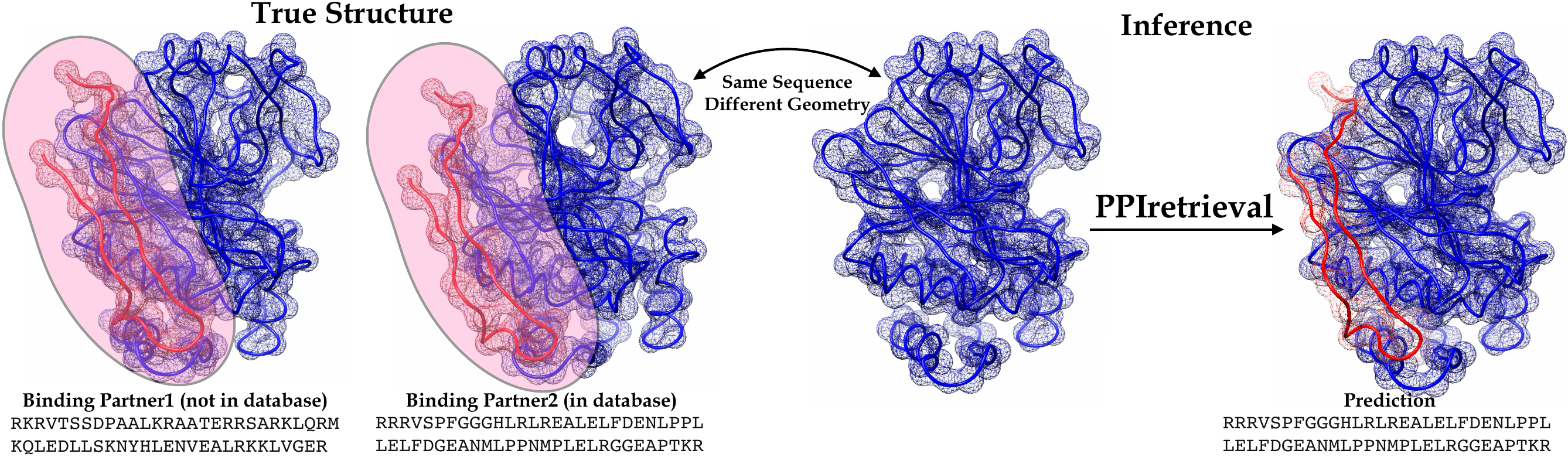}}
\vspace{-0.7cm}
  \caption{
    Case study using PPIretrieval. The query protein, highlighted in blue, successfully identifies a binding partner within our surface database using PPIsearch.
  \vspace{-0.5cm}
  \label{fig:case.study}
}
\end{figure}

\vspace{-0.2cm}
We explore the practical application of PPIretrieval in facilitating the discovery of novel PPIs. 
In real-world scenarios, a protein can have a binding site that interacts with multiple partners. 
While some of these binding partners may be unknown, others might already be known and stored in our database. PPIretrieval can be effectively used to identify these binding partners by finding similar surface representations within the database. A case study demonstrating this application is visualized in Fig.~\ref{fig:case.study}.

In our case study, the query protein has two binding partners: one is already stored in our surface database (\textit{pdb id: 5J28}), while the other is not (\textit{pdb id: 1DGC}). It is important to note that, although the query protein in the two ground-truth structures shares the same sequence representation, there are slight differences in their geometric configuration.

Given the query protein, PPIretrieval identifies the protein in the database that most closely matches in both sequential and geometric representation. Thus, it successfully predicts the corresponding binding partner for this query protein.

\vspace{-0.3cm}
\section{Future Work}
\vspace{-0.2cm}
\label{sec:future.work}
\noindent \textbf{Improving Model Size and Training} \quad
A potential work is to train a larger model with increased parameters to better approximate the protein surface manifold.
Also, we can explore larger models with more hidden dimensions and incorporating more propagation and cross-attention layers.


\vspace{-0.2cm}
\textbf{Dataset Development} \quad
We aim to continuously integrate more high-quality PPI data into the collection of protein complex surface representations. As demonstrated in Tab.~\ref{tab:experiment.cross.dataset}, PPIretrieval exhibits the ability to generalize to unseen proteins. Therefore, future efforts involve training our existing PPIretrieval model on new PPI data, enabling direct encoding and storage of these data in our existing database.

\newpage
\section*{Impact Statements}
PPIretrieval presents work whose goal is to advance the field of AI in drug discovery by designing and finding protein binders, which can potentially improve our understanding of protein complex interactions. We encourage the exploration of novel PPIs findings using PPIretrieval, as it offers significant potential and high impact in this field.

Our goal is to establish a webserver for PPIretrieval, following the service of Foldseek \citep{van2023fast}. As a retrieval model, our objective is to offer a convenient service for individuals interested in exploring novel PPIs using PPIretrieval.

\bibliography{example_paper}
\bibliographystyle{icml2024}

\newpage
\appendix
\onecolumn

\section{Protein-Protein Interaction Visualization}
We visualize additional predicted PPIs carried out by PPIretrieval.
\label{appen:surface.visualization}
\begin{figure}[ht!]
\vspace{-0.3cm}
\centering
{
\includegraphics[width=1.07\columnwidth]{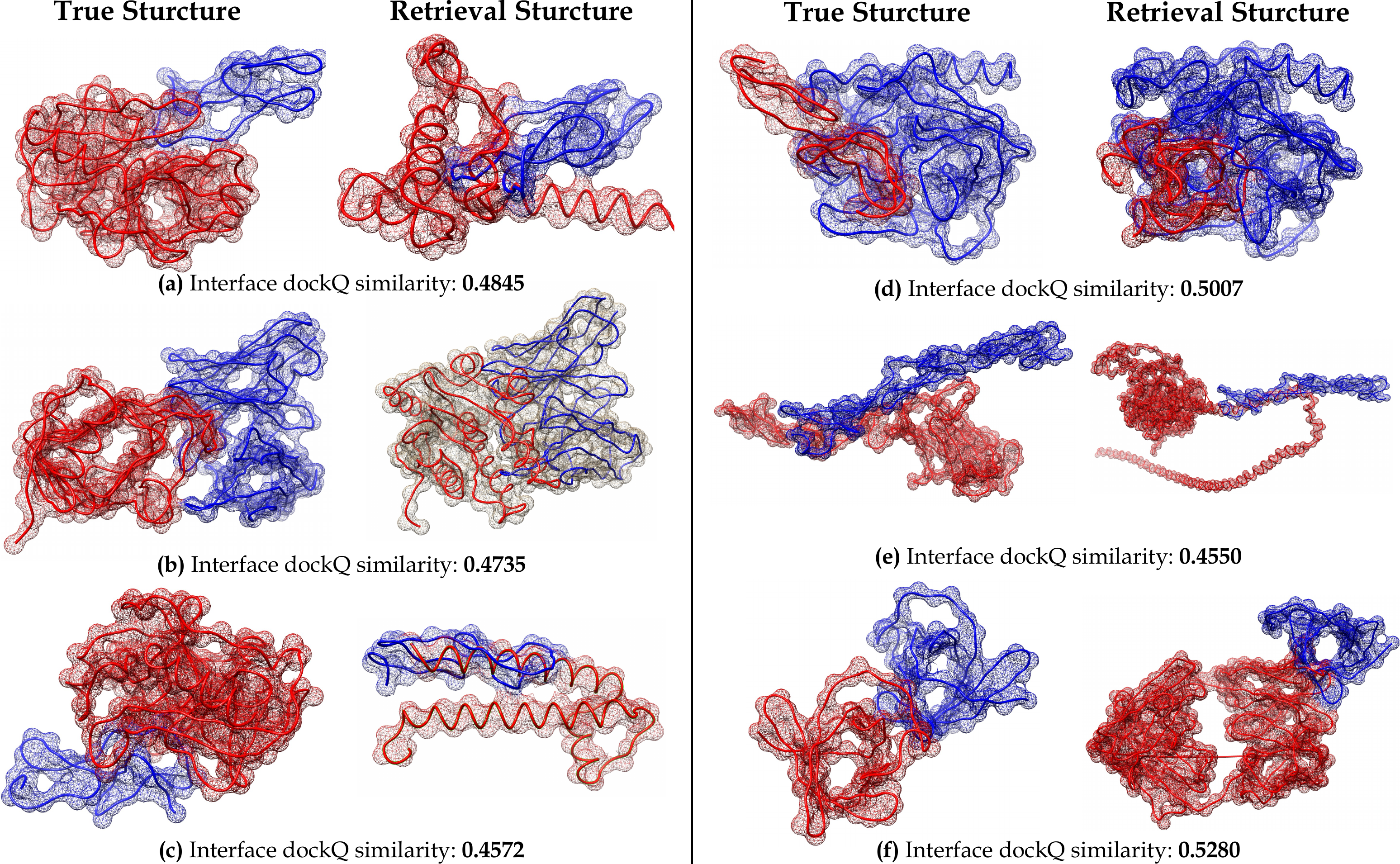}}
\vspace{-0.3cm}
  \caption{Visualization of PPIretrieval results for proteins in the PDB test set, evaluated by \textit{dockQ}. Proteins colored in blue are input query proteins; proteins colored in red are binding partners. Left column displays the ground-truth structures; right column shows the structures predicted by PPIretrieval.
  \vspace{-0.4cm}
  \label{fig:search.visualization.app}
}
\end{figure}

In Fig.~\ref{fig:search.visualization.app}, we observe that the predicted PPIs, with interface \textit{dockQ} similarity, form a well-defined \textit{lock-and-key} structure. This reliable structure formation bolsters confidence in the potential of PPIretrieval for exploring novel protein interactions. One can utilize our model and database to investigate and learn about unknown protein interactions.

\section{Additional Experiments on Empirical Evaluation}
\begin{figure}[ht!]
\centering
{
\includegraphics[width=.5\columnwidth]{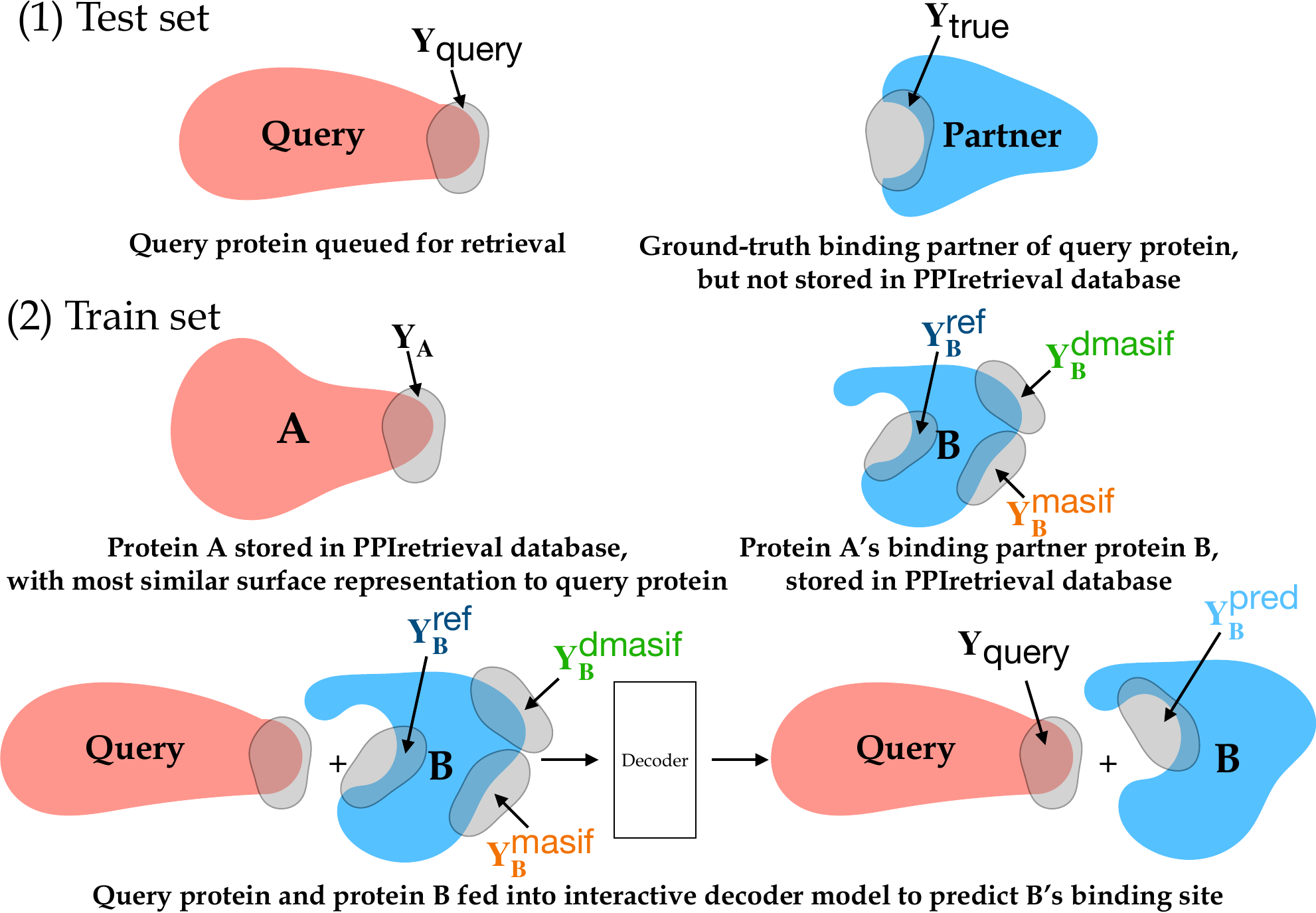}}
  \caption{Evaluation of PPI quality during inference. For a PPI in the test set, a query protein with a known binding site $\mathbf{Y}_\text{query}$ seeks a binding partner with an actual binding site $\mathbf{Y}_\text{true}$. 
  We compute $\textit{dockQ}(\mathbf{Y}_\text{query}, \mathbf{Y}_\text{true}), \textit{TM}(\mathbf{Y}_\text{query}, \mathbf{Y}_\text{true}), \textit{rmsd}(\mathbf{Y}_\text{query}, \mathbf{Y}_\text{true})$ to measure quality of the ground-truth PPI. 
  However, the binding partner is unknown to the PPIretrieval surface database.
  So, PPIretrieval aims to find a potential binding partner from the surface databse.
  PPIretrieval identifies protein $A$ in the surface database, which has the most similar surface representation to the query protein. Protein $A$ has a known binding partner $B$ with a binding site $\mathbf{Y}_B^{\text{ref}}$. 
  We compute $\textit{dockQ}(\mathbf{Y}_\text{query}, \mathbf{Y}_B^{\text{ref}}), \textit{TM}(\mathbf{Y}_\text{query}, \mathbf{Y}_B^{\text{ref}}), \textit{rmsd}(\mathbf{Y}_\text{query}, \mathbf{Y}_B^{\text{ref}})$ to assess quality of the PPI, and $\textit{TM}(\mathbf{Y}_\text{true}, \mathbf{Y}_B^{\text{ref}}), \textit{rmsd}(\mathbf{Y}_\text{true}, \mathbf{Y}_B^{\text{ref}})$ to evaluate quality of the binding sites. 
  PPIretrieval takes query protein and $B$ as input and predicts a new binding site $\mathbf{Y}_B^{\text{pred}}$.  We compute $\textit{dockQ}(\mathbf{Y}_\text{query}, \mathbf{Y}_B^{\text{pred}}), \textit{TM}(\mathbf{Y}_\text{query}, \mathbf{Y}_B^{\text{pred}}), \textit{rmsd}(\mathbf{Y}_\text{query}, \mathbf{Y}_B^{\text{pred}})$ to assess quality of the PPI, 
    $\textit{dockQ}(\mathbf{Y}_\text{true}, \mathbf{Y}_B^{\text{pred}}), \textit{TM}(\mathbf{Y}_\text{true}, \mathbf{Y}_B^{\text{pred}}), \textit{rmsd}(\mathbf{Y}_\text{true}, \mathbf{Y}_B^{\text{pred}})$, $\textit{dockQ}(\mathbf{Y}_\text{true}, \mathbf{Y}_B^{\text{ref}}), \textit{TM}(\mathbf{Y}_\text{true}, \mathbf{Y}_B^{\text{ref}}), \textit{rmsd}(\mathbf{Y}_\text{true}, \mathbf{Y}_B^{\text{ref}})$, $\textit{dockQ}(\mathbf{Y}_\text{true}, \mathbf{Y}_B^{\text{masif}}), \textit{TM}(\mathbf{Y}_\text{true}, \mathbf{Y}_B^{\text{masif}}), \textit{rmsd}(\mathbf{Y}_\text{true}, \mathbf{Y}_B^{\text{masif}})$ to evaluate and compare quality of the binding sites.
  \label{fig:emperical.eval2}
}
\end{figure}

In additional to $\textit{dockQ}(\mathbf{Y}_\text{true}, \mathbf{Y}_B^{\text{pred}}), \textit{TM}(\mathbf{Y}_\text{true}, \mathbf{Y}_B^{\text{pred}}), \textit{rmsd}(\mathbf{Y}_\text{true}, \mathbf{Y}_B^{\text{pred}})$, $\textit{dockQ}(\mathbf{Y}_\text{true}, \mathbf{Y}_B^{\text{ref}}), \textit{TM}(\mathbf{Y}_\text{true}, \mathbf{Y}_B^{\text{ref}}), \textit{rmsd}(\mathbf{Y}_\text{true}, \mathbf{Y}_B^{\text{ref}})$, $\textit{dockQ}(\mathbf{Y}_\text{true}, \mathbf{Y}_B^{\text{masif}}), \textit{TM}(\mathbf{Y}_\text{true}, \mathbf{Y}_B^{\text{masif}}), \textit{rmsd}(\mathbf{Y}_\text{true}, \mathbf{Y}_B^{\text{masif}})$ used in Sec~\ref{sec:experiment}, we evaluate $\textit{dockQ}(\mathbf{Y}_\text{query}, \mathbf{Y}_\text{true}), \textit{TM}(\mathbf{Y}_\text{query}, \mathbf{Y}_\text{true}), \textit{rmsd}(\mathbf{Y}_\text{query}, \mathbf{Y}_\text{true})$, $\textit{dockQ}(\mathbf{Y}_\text{query}, \mathbf{Y}_B^{\text{ref}}), \textit{TM}(\mathbf{Y}_\text{query}, \mathbf{Y}_B^{\text{ref}}), \textit{rmsd}(\mathbf{Y}_\text{query}, \mathbf{Y}_B^{\text{ref}})$,  $\textit{dockQ}(\mathbf{Y}_\text{query}, \mathbf{Y}_B^{\text{pred}}), \textit{TM}(\mathbf{Y}_\text{query}, \mathbf{Y}_B^{\text{pred}}), \textit{rmsd}(\mathbf{Y}_\text{query}, \mathbf{Y}_B^{\text{pred}})$as demonstrated in Fig.~\ref{fig:emperical.eval2}

In Tab.~\ref{tab:ppisearch.baseline.app}, we assess the quality of PPIs and binding sites identified by PPIretrieval with smaller databases. The database for each test set comprises surface representations from the training set of the respective dataset. For example, when evaluating the PDB test set, we only search for surface representations in the PDB training set. Also, the models are trained on each respective dataset.
\begin{table}[htbp!]
\footnotesize
  \centering
  \resizebox{.5\columnwidth}{!}{%
\begin{tabular}{c|cc|c|c|c}
\hline
\hline
\rowcolor[rgb]{ .851,  .851,  .851} \textbf{Dataset} & \multicolumn{2}{c|}{\textbf{Metrics}} & \textbf{PDB} & \textbf{DIPS} & \textbf{PPBS} \\ 
\hline
\multirow{9}{*}{PPI Quality}          & \multicolumn{1}{c|}{\multirow{3}{*}{\textit{dockQ}$(\uparrow)$}} & \cellcolor[rgb]{ 1,  .949,  .8} $\mathbf{Y}_\text{query}, \mathbf{Y}_\text{true}$ &  $0.4596$   &  $0.4797$    &  $0.4949$    \\
& \multicolumn{1}{c|}{} & \cellcolor[rgb]{ 1,  .902,  .6}$\mathbf{Y}_\text{query}, \mathbf{Y}^\text{pred}_B$  &  $\underline{0.4039}$   &  $\underline{0.4042}$    &  $\underline{0.4092}$     \\
& \multicolumn{1}{c|}{} & \cellcolor[rgb]{ .988,  .894,  .839}$\mathbf{Y}_\text{query}, \mathbf{Y}^\text{ref}_B$ &  $0.3907$   &  $0.3935$    &  $0.4040$    \\ \cline{2-6} 
& \multicolumn{1}{c|}{\multirow{3}{*}{\textit{TM}$(\uparrow)$}}    &  \cellcolor[rgb]{ 1,  .949,  .8}$\mathbf{Y}_\text{query}, \mathbf{Y}_\text{true}$ &  $0.4552$   &  $0.5909$    &   $0.5767$    \\
& \multicolumn{1}{c|}{} & \cellcolor[rgb]{ 1,  .902,  .6}$\mathbf{Y}_\text{query}, \mathbf{Y}^\text{pred}_B$ & $\underline{0.2196}$    &  ${0.4211}$    &  $\underline{0.3167}$    \\
& \multicolumn{1}{c|}{}                       & \cellcolor[rgb]{ .988,  .894,  .839}$\mathbf{Y}_\text{query}, \mathbf{Y}^\text{ref}_B$ & $0.1950$    &  $\underline{0.4346}$    &  ${0.3105}$    \\ \cline{2-6} 
& \multicolumn{1}{c|}{\multirow{3}{*}{}}  & \cellcolor[rgb]{ 1,  .949,  .8} $\mathbf{Y}_\text{query}, \mathbf{Y}_\text{true}$ & $7.38$  &  $6.70$    &  $6.70$    \\
& \multicolumn{1}{c|}{\textit{rmsd}$(\downarrow)$} & \cellcolor[rgb]{ 1,  .902,  .6} $\mathbf{Y}_\text{query}, \mathbf{Y}^\text{pred}_B$ & $11.60$    &  $\underline{8.99}$    &  $11.07$    \\
& \multicolumn{1}{c|}{}  & \cellcolor[rgb]{ .988,  .894,  .839}$\mathbf{Y}_\text{query}, \mathbf{Y}^\text{ref}_B$ &  $\underline{10.80}$   &  $10.34$    &   $\underline{10.27}$   \\ 
\hline
\multirow{6}{*}{Site Quality} & \multicolumn{1}{c|}{\multirow{2}{*}{\textit{dockQ}$(\uparrow)$}} & \cellcolor[rgb]{ .851,  .882, .949} $\mathbf{Y}_\text{true}, \mathbf{Y}^\text{pred}_B$  &  $\underline{0.4220}$   & $\underline{0.4304}$     &   $\underline{0.5946}$    \\
& \multicolumn{1}{c|}{}& \cellcolor[rgb]{ .706,  .776,  .906} $\mathbf{Y}_\text{true}, \mathbf{Y}^\text{ref}_B$ &  $0.4073$   &  $0.4177$    &   $0.5535$   \\ 
\cline{2-6} 
& \multicolumn{1}{c|}{\multirow{2}{*}{\textit{TM}$(\uparrow)$}}& \cellcolor[rgb]{ .851,  .882, .949}$\mathbf{Y}_\text{true}, \mathbf{Y}^\text{pred}_B$ &  $\underline{0.2366}$   &  $\underline{0.6649}$    &   $\underline{0.4735}$   \\
& \multicolumn{1}{c|}{}& \cellcolor[rgb]{ .706,  .776,  .906}$\mathbf{Y}_\text{true}, \mathbf{Y}^\text{ref}_B$ &  ${0.2134}$   &  $0.6617$     &  ${0.4622}$     \\
\cline{2-6} 
& \multicolumn{1}{c|}{\multirow{2}{*}{\textit{rmsd}$(\downarrow)$}}& \cellcolor[rgb]{ .851,  .882, .949}$\mathbf{Y}_\text{true}, \mathbf{Y}^\text{pred}_B$ &  $\underline{10.44}$   &  $\underline{6.02}$    &   $9.77$   \\
& \multicolumn{1}{c|}{}& \cellcolor[rgb]{ .706,  .776,  .906}$\mathbf{Y}_\text{true}, \mathbf{Y}^\text{ref}_B$ &  ${11.40}$   &  $11.33$     &  $\underline{8.20}$     \\ 
\hline
\hline
\end{tabular}
}
    \caption{\textit{dockQ}, \textit{TM}, and \textit{rmsd} for evaluation of PPIs and binding sites of \textbf{Top1 hit} predicted by PPIretrieval in comparison with ground-truth structures over three runs. The database for each test set comprises surface representations from the training and validation sets of each respective dataset.}
    \vspace{-0.5cm}
  \label{tab:ppisearch.baseline.app}%
\end{table}

\begin{table}[htbp!]
\footnotesize
  \centering
  \resizebox{.5\columnwidth}{!}{%
\begin{tabular}{c|cc|c|c|c}
\hline
\hline
\rowcolor[rgb]{ .851,  .851,  .851} \textbf{Dataset} & \multicolumn{2}{c|}{\textbf{Metrics}} & \textbf{PDB} & \textbf{DIPS} & \textbf{PPBS} \\ \hline
\multirow{9}{*}{PPI Quality}          & \multicolumn{1}{c|}{\multirow{3}{*}{\textit{dockQ}$(\uparrow)$}} & \cellcolor[rgb]{ 1,  .949,  .8} $\mathbf{Y}_\text{query}, \mathbf{Y}_\text{true}$ &  $0.4596$   &  $0.4797$   &  $0.4949$    \\
& \multicolumn{1}{c|}{} & \cellcolor[rgb]{ 1,  .902,  .6}$\mathbf{Y}_\text{query}, \mathbf{Y}^\text{pred}_B$  &  $\underline{0.4110}$   &  $\underline{0.4394}$    &  $\underline{0.4400}$    \\
& \multicolumn{1}{c|}{} & \cellcolor[rgb]{ .988,  .894,  .839}$\mathbf{Y}_\text{query}, \mathbf{Y}^\text{ref}_B$ & $0.4061$    &  $0.4093$    &   $0.4130$   \\ \cline{2-6} 
& \multicolumn{1}{c|}{\multirow{3}{*}{\textit{TM}$(\uparrow)$}}    &  \cellcolor[rgb]{ 1,  .949,  .8}$\mathbf{Y}_\text{query}, \mathbf{Y}_\text{true}$ & $0.4552$    &   $0.5909$   &  $0.5767$    \\
& \multicolumn{1}{c|}{} & \cellcolor[rgb]{ 1,  .902,  .6}$\mathbf{Y}_\text{query}, \mathbf{Y}^\text{pred}_B$ &  $\underline{0.2649}$   &   $\underline{0.4507}$   &  $\underline{0.4148}$    \\
& \multicolumn{1}{c|}{}                       & \cellcolor[rgb]{ .988,  .894,  .839}$\mathbf{Y}_\text{query}, \mathbf{Y}^\text{ref}_B$ &  ${0.2507}$   &  $0.4394$    &   $0.4009$   \\ \cline{2-6} 
& \multicolumn{1}{c|}{\multirow{3}{*}{}}  & \cellcolor[rgb]{ 1,  .949,  .8} $\mathbf{Y}_\text{query}, \mathbf{Y}_\text{true}$ &  $7.38$   &   $6.70$   &  $6.70$    \\
& \multicolumn{1}{c|}{\textit{rmsd}$(\downarrow)$} & \cellcolor[rgb]{ 1,  .902,  .6} $\mathbf{Y}_\text{query}, \mathbf{Y}^\text{pred}_B$ & $10.78$    &  $\underline{8.84}$    &  $\underline{9.11}$    \\
& \multicolumn{1}{c|}{}  & \cellcolor[rgb]{ .988,  .894,  .839}$\mathbf{Y}_\text{query}, \mathbf{Y}^\text{ref}_B$ &  $\underline{9.88}$   &  $10.09$    &  $11.01$    \\ \hline
\multirow{6}{*}{Site Quality} & \multicolumn{1}{c|}{\multirow{2}{*}{\textit{dockQ}$(\uparrow)$}} & \cellcolor[rgb]{ .851,  .882, .949} $\mathbf{Y}_\text{true}, \mathbf{Y}^\text{pred}_B$  & $\underline{0.4678}$    &  $\underline{0.4410}$    &  $\underline{0.6045}$    \\
& \multicolumn{1}{c|}{}& \cellcolor[rgb]{ .706,  .776,  .906} $\mathbf{Y}_\text{true}, \mathbf{Y}^\text{ref}_B$ & $0.4250$    &  ${0.4367}$    &  $0.6014$    \\ 
\cline{2-6} 
& \multicolumn{1}{c|}{\multirow{2}{*}{\textit{TM}$(\uparrow)$}} & \cellcolor[rgb]{ .851,  .882, .949} $\mathbf{Y}_\text{true}, \mathbf{Y}^\text{pred}_B$  & $\underline{0.3300}$    &  $\underline{0.6944}$    &  $\underline{0.6045}$    \\
& \multicolumn{1}{c|}{}& \cellcolor[rgb]{ .706,  .776,  .906} $\mathbf{Y}_\text{true}, \mathbf{Y}^\text{ref}_B$ & $0.3222$    &  ${0.6914}$    &  $0.6014$    \\
\cline{2-6} 
& \multicolumn{1}{c|}{\multirow{2}{*}{\textit{rmsd}$(\downarrow)$}}& \cellcolor[rgb]{ .851,  .882, .949}$\mathbf{Y}_\text{true}, \mathbf{Y}^\text{pred}_B$ &  $10.70$   &  $\underline{5.67}$    &   $\underline{6.52}$   \\
& \multicolumn{1}{c|}{}& \cellcolor[rgb]{ .706,  .776,  .906}$\mathbf{Y}_\text{true}, \mathbf{Y}^\text{ref}_B$ & $\underline{9.30}$    &   $9.65$   &  $9.96$    \\ 
\hline
\hline
\end{tabular}
}
    \caption{\textit{dockQ}, \textit{TM}, and \textit{rmsd} for evaluation of PPIs and binding sites of \textbf{Top1 hit} predicted by PPIretrieval in comparison with ground-truth structures over three runs. The database comprises all surface representations from the training and validation sets of PDB, DIPS, and PPBS datasets.}
  \label{tab:ppisearch.baseline2.app}%
\end{table}

\begin{table}[htbp!]
\footnotesize
  \centering
  \resizebox{.5\columnwidth}{!}{%
\begin{tabular}{c|cc|c|c|c|c}
\hline
\hline
\rowcolor[rgb]{ .851,  .851,  .851} \textbf{Dataset} & \multicolumn{2}{c|}{\textbf{Metrics}} & \textbf{DIPS-Top1} & \textbf{DIPS-Top10} & \textbf{PPBS-Top1} & \textbf{PPBS-Top10} \\ \hline
\multirow{9}{*}{PPI Quality}          & \multicolumn{1}{c|}{\multirow{3}{*}{\textit{dockQ}$(\uparrow)$}} & \cellcolor[rgb]{ 1,  .949,  .8} $\mathbf{Y}_\text{query}, \mathbf{Y}_\text{true}$ &  $0.4797$   &  $0.4797$    & $0.4949$  &  $0.4949$ \\
& \multicolumn{1}{c|}{} & \cellcolor[rgb]{ 1,  .902,  .6}$\mathbf{Y}_\text{query}, \mathbf{Y}^\text{pred}_B$  &  $\underline{0.4334}$   &  $\cellcolor[rgb]{ .573,  .816,  .314}\underline{0.4432}$ & $\underline{0.4098}$  & $\cellcolor[rgb]{ .573,  .816,  .314}\underline{0.4447}$  \\
& \multicolumn{1}{c|}{} & \cellcolor[rgb]{ .988,  .894,  .839}$\mathbf{Y}_\text{query}, \mathbf{Y}^\text{ref}_B$ &  $0.4054$   &   $0.4316$   & $0.4040$  & $0.4051$  \\ \cline{2-7} 
& \multicolumn{1}{c|}{\multirow{3}{*}{\textit{TM}$(\uparrow)$}}    &  \cellcolor[rgb]{ 1,  .949,  .8}$\mathbf{Y}_\text{query}, \mathbf{Y}_\text{true}$ &  $0.5909$   &  $0.5909$    & $0.5767$  & $0.5767$  \\
& \multicolumn{1}{c|}{} & \cellcolor[rgb]{ 1,  .902,  .6}$\mathbf{Y}_\text{query}, \mathbf{Y}^\text{pred}_B$ &  $\underline{0.3965}$   &   $\cellcolor[rgb]{ .573,  .816,  .314}\underline{0.4691}$   & $\cellcolor[rgb]{ .573,  .816,  .314}\underline{0.3219}$  & $\underline{0.3021}$  \\
& \multicolumn{1}{c|}{}                       & \cellcolor[rgb]{ .988,  .894,  .839}$\mathbf{Y}_\text{query}, \mathbf{Y}^\text{ref}_B$ &  $0.3708$   &   $0.4357$   &  $0.2903$ &  $0.2780$ \\ \cline{2-7} 
& \multicolumn{1}{c|}{\multirow{3}{*}{}}  & \cellcolor[rgb]{ 1,  .949,  .8} $\mathbf{Y}_\text{query}, \mathbf{Y}_\text{true}$ &  $6.70$   &  $6.70$    & $6.70$ & $6.70$   \\
& \multicolumn{1}{c|}{\textit{rmsd}$(\downarrow)$} & \cellcolor[rgb]{ 1,  .902,  .6} $\mathbf{Y}_\text{query}, \mathbf{Y}^\text{pred}_B$ &  $\underline{8.71}$   &   $10.49$   &  $12.09$ & $10.25$   \\
& \multicolumn{1}{c|}{}  & \cellcolor[rgb]{ .988,  .894,  .839}$\mathbf{Y}_\text{query}, \mathbf{Y}^\text{ref}_B$ &  $11.68$   &  $\underline{9.26}$    & $\underline{10.71}$  & $\underline{10.13}$  \\ \hline
\multirow{6}{*}{Site Quality} & \multicolumn{1}{c|}{\multirow{2}{*}
{\textit{dockQ}$(\uparrow)$}} & \cellcolor[rgb]{ .851,  .882, .949} $\mathbf{Y}_\text{true}, \mathbf{Y}^\text{pred}_B$  &  $\underline{0.4207}$   &  $\underline{0.4435}$    & $\underline{0.5579}$  & $\underline{0.5857}$  \\
& \multicolumn{1}{c|}{}& \cellcolor[rgb]{ .706,  .776,  .906} $\mathbf{Y}_\text{true}, \mathbf{Y}^\text{ref}_B$ & $0.4030$   &  $0.4156$    & $0.5231$ &  $0.5611$  \\ 
\cline{2-7} 
& \multicolumn{1}{c|}{\multirow{2}{*}
{\textit{TM}$(\uparrow)$}} & \cellcolor[rgb]{ .851,  .882, .949} $\mathbf{Y}_\text{true}, \mathbf{Y}^\text{pred}_B$  &  $\underline{0.5419}$   &  $\underline{0.6792}$    & $\underline{0.4421}$  & $\underline{0.3889}$  \\
& \multicolumn{1}{c|}{}& \cellcolor[rgb]{ .706,  .776,  .906} $\mathbf{Y}_\text{true}, \mathbf{Y}^\text{ref}_B$ & $0.5330$   &  $0.6714$    & $0.4202$ &  $0.3725$  \\  
\cline{2-7} 
& \multicolumn{1}{c|}{\multirow{2}{*}{\textit{rmsd}$(\downarrow)$}}& \cellcolor[rgb]{ .851,  .882, .949}$\mathbf{Y}_\text{true}, \mathbf{Y}^\text{pred}_B$ &  $\underline{5.84}$   &  $10.50$    &   $11.76$ & $10.49$ \\
& \multicolumn{1}{c|}{}& \cellcolor[rgb]{ .706,  .776,  .906}$\mathbf{Y}_\text{true}, \mathbf{Y}^\text{ref}_B$ &  $11.12$   &  $\underline{7.35}$    & $\underline{8.92}$  & $\underline{8.91}$  \\ 
\hline
\hline
\end{tabular}
}
    \caption{\textit{dockQ}, \textit{TM}, and \textit{rmsd} for evaluation of PPIs and binding sites of \textbf{Top1, Top10 hit} predicted by PPIretrieval in comparison with ground-truth structures on cross-datasets over three runs. The database for each test set comprises surface representations from the training and validation sets of each respective dataset.}
  \label{tab:experiment.cross.dataset.app}%
\end{table}

\begin{figure}[ht!]
\centering
{\includegraphics[width=.5\columnwidth]{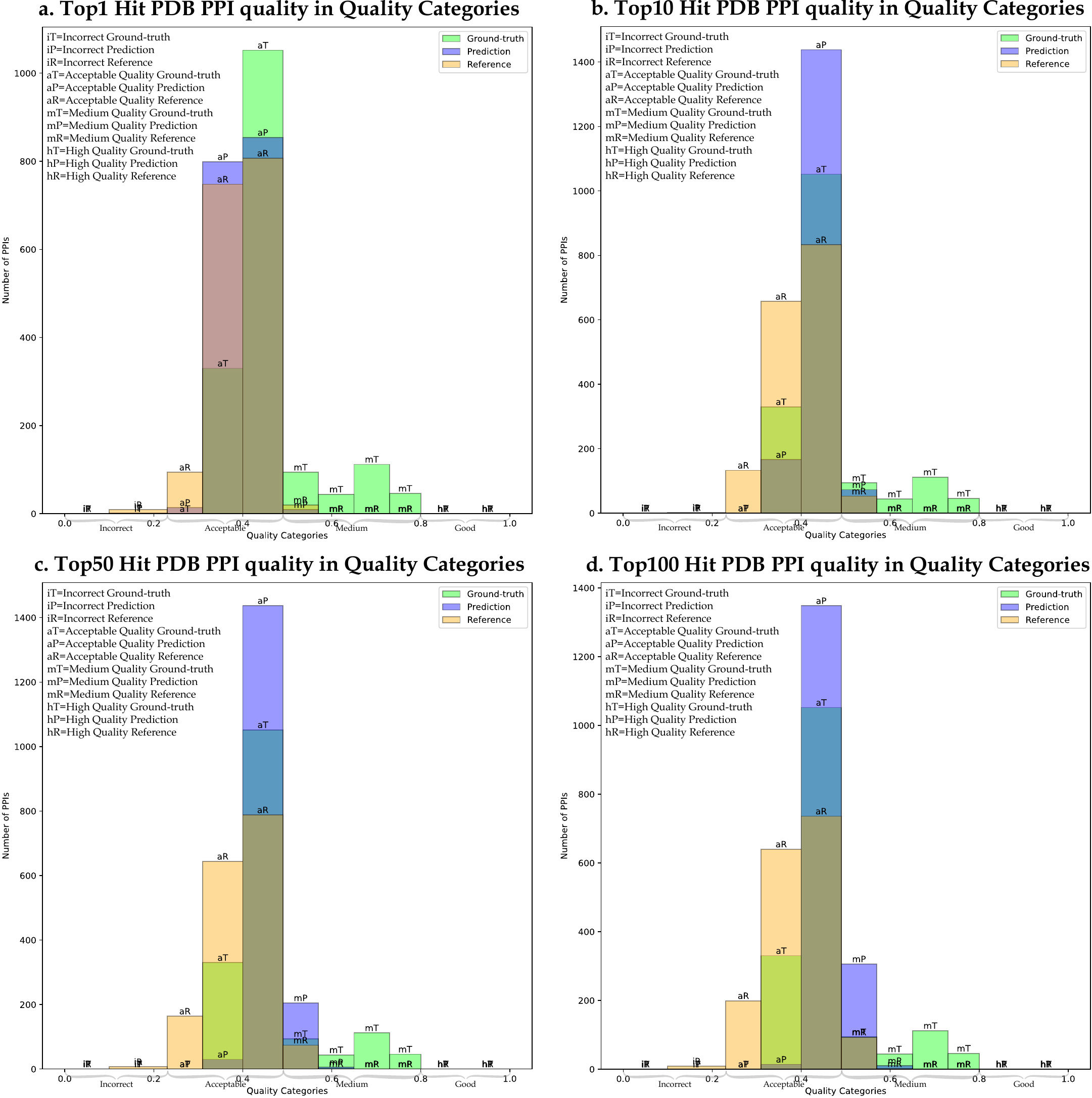}}
  \caption{Comparison of PPI qualities in the test set of PDB dataset, considering ground-truth, predicted, and reference PPIs, evaluated using the \textit{dockQ} score at \textbf{Top1, Top10, Top50, Top100 hit}. The database comprises surface representations from training and validation sets from PDB dataset only.
  \label{fig:quality.category.distribution}
}
\end{figure}

In Tab.~\ref{tab:ppisearch.baseline2.app}, we evaluate the quality of PPIs and binding sites identified by PPIretrieval with larger database. The database includes surface representations from the training and validation sets of the PDB, DIPS, and PPBS datasets, in total of $155,384$ paired proteins with their surface features for retrieval. And the models are trained on all training PPIs.

In Tab.~\ref{tab:experiment.cross.dataset.app}, we show the cross-dataset performance. We take the model trained on PDB training set only to encode the PPIs in DIPS and PPBS training and validation sets, respectively. Then we evaluate the cross-dataset performance on DIPS and PPBS test sets at two different hit rates, respectively. \textbf{Top10 hit} means that we identify the $10$ most similar surface representations to the query protein in the test set for inference. Then, PPIretrieval decodes between the query protein and the $10$ potential binding partners associated with these similar proteins. The best binding partner for the query protein is then selected based on the highest \textit{dockQ} score.

Additionally, we show the quality distribution of PPIs predicted by PPIretrieval in comparison with ground-truth and reference at different hit rates in Fig~\ref{fig:quality.category.distribution}. For \textit{dockQ} score \citep{basu2016dockq}, a score in the range (0, 0.23) denotes incorrect interaction, [0.23, 0.49) denotes acceptable interaction, [0.49, 0.8) denotes medium interaction, and [0.8, 1) denotes good interaction. In Fig~\ref{fig:quality.category.distribution}, we divide these ranges into sub-ranges for better visualization of the quality distribution, (0, 0.23)$\rightarrow$(0, 0.1)$\cup$[0.1, 0.23), [0.23, 0.49)$\rightarrow$[0.23, 0.31)$\cup$[0.31, 0.4)$\cup$[0.4, 0.49), [0.49, 0.8)$\rightarrow$[0.49, 0.57)$\cup$[0.57, 0.65)$\cup$[0.65, 0.73)$\cup$[0.65, 0.73)$\cup$[0.65, 0.8),[0.8, 1)$\rightarrow$[0.8, 0.9)$\cup$[0.9, 1).

We observe that more predicted PPIs fall into the medium-quality category as hit rates increase from \textbf{Top1} to \textbf{Top100}, surpassing the number of ground-truth PPIs of acceptable and medium qualities. This consistent quality distribution of PPIs predicted by PPIretrieval indicates a strong potential for novel PPI findings. Furthermore, we compare the results of using cosine similarity in Tab.~\ref{tab:pdb.euc.vs.cos} and Fig.~\ref{fig:pdb.euc.cos.search}.

\section{Additional Experiments on Euclidean Distance vs. Cosine Similarity}
\label{app:more.experiments.quality.distribution}
We show additional results and visualize more quality distributions carried out by PPIretrieval using Euclidean distance and cosine similarity in this section.

In Sec.~\ref{sec:ppisearch.pipeline.overview} and Fig.~\ref{fig:ppisearch.workflow}, we present the PPIretrieval inference strategy to find a binding partner for a query protein. We identify a similar surface representation to the query protein in the PPIretrieval database using the Euclidean distance function for the surface embeddings,
\begin{equation}
    d(\mathbf{H}_P, \mathbf{H}_A) = \sqrt{\sum_{i=1}^d \left( \frac{1}{N}\sum_{j=1}^N \mathbf{H}_{P}^{j, i} - \frac{1}{M}\sum_{k=1}^M \mathbf{H}_{A}^{k, i} \right)^2} \in \mathbb{R}.
\end{equation}
Here, $\mathbf{H}_P \in \mathbb{R}^{N\times d}$ represents the surface embedding of the query protein, and $\mathbf{H}_A \in \mathbb{R}^{M\times d}$ represents the surface embedding of a protein in our PPIretrieval database. In addition to the Euclidean distance, we can use cosine similarity to find the surface representation that is the most similar to the query protein.

In Tab.~\ref{tab:pdb.euc.vs.cos}, we present experimental results for PPIretrieval using Euclidean Distance and Cosine Similarity at five different hit rates, ranging from \textbf{Top1} to \textbf{Top100}. The models are exclusively trained on the PDB training set. Our observations reveal an improvement in the quality of predicted PPIs, as measured by the \textit{dockQ} score and \textit{rmsd}, when employing Cosine Similarity to find similar surface representations compared to using Euclidean Distance for the same purpose in the PPIretrieval database. During the inference stage, we leverage Cosine Similarity as an alternative method for retrieving similar surface representations for the query protein, in contrast to the original approach using Euclidean distance.
 
\begin{table}[ht!]
    \begin{minipage}{.5\linewidth}\centering
        \resizebox{.9\textwidth}{!}{%
\begin{tabular}{c|cc|c|c|c|c|c}
\hline
\hline
\rowcolor[rgb]{ .851,  .851,  .851} \textbf{PDB Dataset} & \multicolumn{2}{c|}{\textbf{Metrics}} & \textbf{Top1} & \textbf{Top10} & \textbf{Top20} & \textbf{Top50} & \textbf{Top100} \\ \hline
\multirow{9}{*}{PPI Quality}          & \multicolumn{1}{c|}{\multirow{3}{*}{\textit{dockQ}$(\uparrow)$}} & \cellcolor[rgb]{ 1,  .949,  .8} $\mathbf{Y}_\text{query}, \mathbf{Y}_\text{true}$ &  $0.4596$   & $0.4596$  & $0.4596$ & $0.4596$  &  $0.4596$    \\
& \multicolumn{1}{c|}{} & \cellcolor[rgb]{ 1,  .902,  .6}$\mathbf{Y}_\text{query}, \mathbf{Y}^\text{pred}_B$  &  $\underline{0.4039}$   &  $\underline{0.4352}$    & $\underline{0.4428}$  & $\underline{0.4532}$  &  \cellcolor[rgb]{ .573,  .816,  .314}$\underline{0.4614}$   \\
& \multicolumn{1}{c|}{} & \cellcolor[rgb]{ .988,  .894,  .839}$\mathbf{Y}_\text{query}, \mathbf{Y}^\text{ref}_B$ &  $0.3907$   &  $0.3920$  & $0.3943$  & $0.3963$ & $0.3968$  \\ \cline{2-8} 
& \multicolumn{1}{c|}{\multirow{3}{*}{\textit{TM}$(\uparrow)$}}    &  \cellcolor[rgb]{ 1,  .949,  .8}$\mathbf{Y}_\text{query}, \mathbf{Y}_\text{true}$ &  $0.4552$   &   $0.4552$   & $0.4552$ & $0.4552$ & $0.4552$   \\
& \multicolumn{1}{c|}{} & \cellcolor[rgb]{ 1,  .902,  .6}$\mathbf{Y}_\text{query}, \mathbf{Y}^\text{pred}_B$ & $\underline{0.2196}$    &  $\underline{0.2183}$    & $\underline{0.0.2174}$  & $\underline{0.2168}$ & $\underline{0.2159}$  \\
& \multicolumn{1}{c|}{}                       & \cellcolor[rgb]{ .988,  .894,  .839}$\mathbf{Y}_\text{query}, \mathbf{Y}^\text{ref}_B$ & $0.1950$    &  $0.1938$  & $0.1944$  & $0.1950$ & $0.1958$  \\ \cline{2-8} 
& \multicolumn{1}{c|}{\multirow{3}{*}{}}  & \cellcolor[rgb]{ 1,  .949,  .8} $\mathbf{Y}_\text{query}, \mathbf{Y}_\text{true}$ & $7.38$  & $7.38$    &  $7.38$ & $7.38$ &  $7.38$ \\
& \multicolumn{1}{c|}{\textit{rmsd}$(\downarrow)$} & \cellcolor[rgb]{ 1,  .902,  .6} $\mathbf{Y}_\text{query}, \mathbf{Y}^\text{pred}_B$ & $11.60$    &  $\underline{9.52}$    & $\underline{9.17}$  &  $\underline{8.73}$ & \cellcolor[rgb]{ .573,  .816,  .314}$\underline{8.42}$  \\
& \multicolumn{1}{c|}{}  & \cellcolor[rgb]{ .988,  .894,  .839}$\mathbf{Y}_\text{query}, \mathbf{Y}^\text{ref}_B$ &  $\underline{10.80}$   & $9.77$  & $9.65$  & $9.65$ &  $9.56$ \\ \hline
\multirow{4}{*}{Site Quality} & \multicolumn{1}{c|}{\multirow{2}{*}{\textit{TM}$(\uparrow)$}} & \cellcolor[rgb]{ .851,  .882, .949} $\mathbf{Y}_\text{true}, \mathbf{Y}^\text{pred}_B$  &  $\underline{0.2366}$   &  $\underline{0.2195}$ & $\underline{0.2174}$ & $\underline{0.2154}$ &  $\underline{0.2156}$  \\
& \multicolumn{1}{c|}{}& \cellcolor[rgb]{ .706,  .776,  .906} $\mathbf{Y}_\text{true}, \mathbf{Y}^\text{ref}_B$ &  $0.2134$   &  $0.1986$  & $0.1960$  & $0.1955$ & $0.1957$  \\ \cline{2-8} 
& \multicolumn{1}{c|}{\multirow{2}{*}{\textit{rmsd}$(\downarrow)$}}& \cellcolor[rgb]{ .851,  .882, .949}$\mathbf{Y}_\text{true}, \mathbf{Y}^\text{pred}_B$ &  $11.52$   &  $10.01$    & $9.84$ & $9.62$ &  $\cellcolor[rgb]{ .573,  .816,  .314}\underline{9.42}$  \\
& \multicolumn{1}{c|}{}& \cellcolor[rgb]{ .706,  .776,  .906}$\mathbf{Y}_\text{true}, \mathbf{Y}^\text{ref}_B$ &  $\underline{10.50}$   &  $\underline{9.69}$    & $\underline{9.62}$  & $\underline{9.53}$ & $9.47$  \\ 
\hline
\hline
\end{tabular}
}
\caption{Retrieval Using Euclidean Distance}
    \end{minipage}%
    \begin{minipage}{.5\linewidth}\centering
        \resizebox{.9\textwidth}{!}{%
\begin{tabular}{c|cc|c|c|c|c|c}
\hline
\hline
\rowcolor[rgb]{ .851,  .851,  .851} \textbf{PDB Dataset} & \multicolumn{2}{c|}{\textbf{Metrics}} & \textbf{Top1} & \textbf{Top10} & \textbf{Top20} & \textbf{Top50} & \textbf{Top100} \\ \hline
\multirow{9}{*}{PPI Quality}          & \multicolumn{1}{c|}{\multirow{3}{*}{\textit{dockQ}$(\uparrow)$}} & \cellcolor[rgb]{ 1,  .949,  .8} $\mathbf{Y}_\text{query}, \mathbf{Y}_\text{true}$ &  $0.4596$   & $0.4596$  & $0.4596$ & $0.4596$  &  $0.4596$    \\
& \multicolumn{1}{c|}{} & \cellcolor[rgb]{ 1,  .902,  .6}$\mathbf{Y}_\text{query}, \mathbf{Y}^\text{pred}_B$  &  $\underline{0.3907}$   &  $\underline{0.4381}$    & $\underline{0.4472}$  & $\underline{0.4601}$  &  \cellcolor[rgb]{ .573,  .9,  .314}$\underline{0.4683}$   \\
& \multicolumn{1}{c|}{} & \cellcolor[rgb]{ .988,  .894,  .839}$\mathbf{Y}_\text{query}, \mathbf{Y}^\text{ref}_B$ &  $0.3147$   &  $0.3319$  & $0.3388$  & $0.3407$ & $0.3422$  \\ \cline{2-8} 
& \multicolumn{1}{c|}{\multirow{3}{*}{\textit{TM}$(\uparrow)$}}    &  \cellcolor[rgb]{ 1,  .949,  .8}$\mathbf{Y}_\text{query}, \mathbf{Y}_\text{true}$ &  $0.4552$   &   $0.4552$   & $0.4552$ & $0.4552$ & $0.4552$   \\
& \multicolumn{1}{c|}{} & \cellcolor[rgb]{ 1,  .902,  .6}$\mathbf{Y}_\text{query}, \mathbf{Y}^\text{pred}_B$ & $\underline{0.2009}$    &  $\underline{0.2101}$    & $\underline{0.2085}$  & $\underline{0.2089}$ & $\underline{0.2095}$  \\
& \multicolumn{1}{c|}{}                       & \cellcolor[rgb]{ .988,  .894,  .839}$\mathbf{Y}_\text{query}, \mathbf{Y}^\text{ref}_B$ & $0.1838$    &  $0.1857$  & $0.1861$  & $0.1934$ & $0.1941$  \\ \cline{2-8} 
& \multicolumn{1}{c|}{\multirow{3}{*}{}}  & \cellcolor[rgb]{ 1,  .949,  .8} $\mathbf{Y}_\text{query}, \mathbf{Y}_\text{true}$ & $7.38$  & $7.38$    &  $7.38$ & $7.38$ &  $7.38$ \\
& \multicolumn{1}{c|}{\textit{rmsd}$(\downarrow)$} & \cellcolor[rgb]{ 1,  .902,  .6} $\mathbf{Y}_\text{query}, \mathbf{Y}^\text{pred}_B$ & $11.08$    &  $\underline{9.15}$    & $\underline{8.86}$  &  $\underline{8.39}$ & \cellcolor[rgb]{ .573,  .9,  .314}$\underline{7.99}$  \\
& \multicolumn{1}{c|}{}  & \cellcolor[rgb]{ .988,  .894,  .839}$\mathbf{Y}_\text{query}, \mathbf{Y}^\text{ref}_B$ &  $\underline{11.01}$   & $9.91$  & $9.81$  & $9.92$ &  $9.79$ \\ \hline
\multirow{4}{*}{Site Quality} & \multicolumn{1}{c|}{\multirow{2}{*}{\textit{TM}$(\uparrow)$}} & \cellcolor[rgb]{ .851,  .882, .949} $\mathbf{Y}_\text{true}, \mathbf{Y}^\text{pred}_B$  &  $\underline{0.1981}$   &  $\underline{0.2084}$ & $\underline{0.2082}$ & $\underline{0.2049}$ &  $\underline{0.2055}$  \\
& \multicolumn{1}{c|}{}& \cellcolor[rgb]{ .706,  .776,  .906} $\mathbf{Y}_\text{true}, \mathbf{Y}^\text{ref}_B$ &  $0.1818$   &  $0.1832$  & $0.1860$  & $0.1923$ & $0.1919$  \\ \cline{2-8} 
& \multicolumn{1}{c|}{\multirow{2}{*}{\textit{rmsd}$(\downarrow)$}}& \cellcolor[rgb]{ .851,  .882, .949}$\mathbf{Y}_\text{true}, \mathbf{Y}^\text{pred}_B$ &  $11.05$   &  $9.56$    & $9.42$ & $9.13$ &  $\cellcolor[rgb]{ .573,  .9,  .314}\underline{9.09}$  \\
& \multicolumn{1}{c|}{}& \cellcolor[rgb]{ .706,  .776,  .906}$\mathbf{Y}_\text{true}, \mathbf{Y}^\text{ref}_B$ &  $\underline{10.90}$   &  $\underline{9.93}$    & $\underline{9.83}$  & $\underline{9.80}$ & $9.87$  \\ 
\hline
\hline
\end{tabular}
}
\caption{Retrieval Using Consine Similarity}
    \end{minipage} 
    \caption{\textit{dockQ}, \textit{TM}, and \textit{rmsd} for evaluation of PPIs and binding sites of \textbf{Top1, Top10, Top20, Top50, Top100 hit} predicted by PPIretrieval in comparison with ground-truth structures on PDB dataset over three runs. The database comprises surface representations from training and validation sets from PDB dataset only.}
    \label{tab:pdb.euc.vs.cos}
\end{table}

\begin{figure}[ht!]
\subfigure[Retrieval Using Euclidean Distance]{{\includegraphics[width=0.5\columnwidth]{PDB_quality_distribution.pdf}}}
\subfigure[Retrieval Using Cosine Similarity]{{\includegraphics[width=0.5\columnwidth]{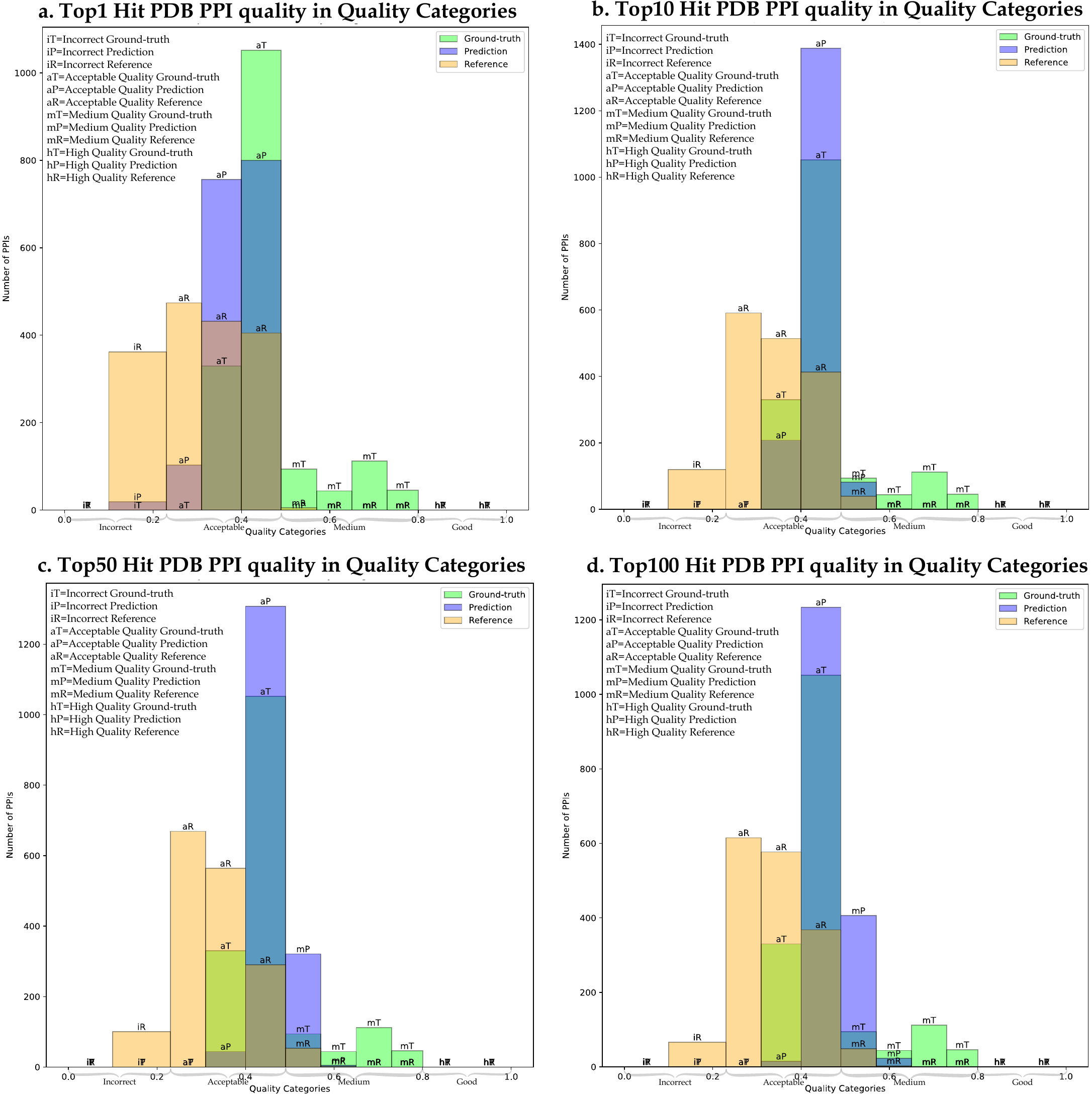}}}
\caption{Comparison of PPI qualities in the test set of PDB dataset, considering ground-truth, predicted, and reference PPIs, evaluated using the \textit{dockQ} score at \textbf{Top1, Top10, Top50, Top100 hit}. The database comprises surface representations from training and validation sets from PDB dataset only.}
\label{fig:pdb.euc.cos.search}
\end{figure}

In addition to the tabular results in Tab.~\ref{tab:pdb.euc.vs.cos}, we visualize the quality distribution of predicted PPIs by PPIretrieval using Euclidean Distance and Cosine Similarity in Fig.~\ref{fig:pdb.euc.cos.search}. Our observations reveal that, as hit rates increase from \textbf{Top1} to \textbf{Top100}, more predicted PPIs fall into the medium-quality category when using cosine similarity to retrieve surface representations similar to the query protein. This trend exceeds the number of ground-truth PPIs of acceptable and medium qualities. The consistent quality distribution of predicted PPIs by PPIretrieval suggests a robust potential for novel PPI exploration with PPIretrieval. Additionally, we provide Cosine Similarity as an alternative choice to the Euclidean distance in our approach.

\section{Additional Experiments on Cross-Dataset Validation}
We show additional results and visualize more quality distributions carried out by PPIretrieval on the cross-dataset validation. Following the cross-dataset validation results in Tab.~\ref{tab:experiment.cross.dataset}, we provide a thorough experimental analysis and visualization here.

In Table~\ref{tab:pdb.vs.ppbs.crossdataset}, we present the cross-dataset validation results for \textbf{Top1} and \textbf{Top10 hits} on the PPBS test set. We utilize two models for this analysis: one is trained on the PPBS training set and validated on the PPBS test set, and the other is trained on the PDB training set and \textit{cross-validated} on the PPBS test set. With \textbf{Top1 hit}, we find that PPIretrieval is capable of generalizing to unseen protein complexes, as evidenced by the second model exhibiting better PPI quality compared to the first model in terms of \textit{dockQ} and \textit{TM} scores. When we expand our retrieval space to \textbf{Top10 hit}, the first model (trained on the PPBS training set) predicts PPIs with improved quality.

\begin{table}[htbp!]
\footnotesize
  \centering
  \resizebox{0.8\columnwidth}{!}{%
\begin{tabular}{c|cc|c|c|c|c}
\hline
\hline
\rowcolor[rgb]{ .851,  .851,  .851} \textbf{Dataset} & \multicolumn{2}{c|}{\textbf{Metrics}} & \textbf{PPBS-Top1 } & \textbf{PPBS-Top1 (PDB Cross-Dataset)} & \textbf{PPBS-Top10 } & \textbf{PPBS-Top10 (PDB Cross-Dataset)} \\ \hline
\multirow{9}{*}{PPI Quality}          & \multicolumn{1}{c|}{\multirow{3}{*}{\textit{dockQ}$(\uparrow)$}} & \cellcolor[rgb]{ 1,  .949,  .8} $\mathbf{Y}_\text{query}, \mathbf{Y}_\text{true}$ &  $0.4949$   &  $0.4949$    & $0.4949$  &  $0.4949$ \\
& \multicolumn{1}{c|}{} & \cellcolor[rgb]{ 1,  .902,  .6}$\mathbf{Y}_\text{query}, \mathbf{Y}^\text{pred}_B$  &  ${0.4092}$   &  $\underline{0.4098}$ & $\cellcolor[rgb]{ .573,  .816,  .314}\underline{0.4513}$  & $0.4447$  \\
& \multicolumn{1}{c|}{} & \cellcolor[rgb]{ .988,  .894,  .839}$\mathbf{Y}_\text{query}, \mathbf{Y}^\text{ref}_B$ &  $0.4040$   &   $0.4040$   & $0.4345$  & $0.4051$  \\ \cline{2-7} 
& \multicolumn{1}{c|}{\multirow{3}{*}{\textit{TM}$(\uparrow)$}}    &  \cellcolor[rgb]{ 1,  .949,  .8}$\mathbf{Y}_\text{query}, \mathbf{Y}_\text{true}$ &  $0.5767$   &  $0.5767$    & $0.5767$  & $0.5767$  \\
& \multicolumn{1}{c|}{} & \cellcolor[rgb]{ 1,  .902,  .6}$\mathbf{Y}_\text{query}, \mathbf{Y}^\text{pred}_B$ &  ${0.3167}$   &   $\underline{0.3219}$   & $\cellcolor[rgb]{ .573,  .816,  .314}\underline{0.3245}$  & ${0.3021}$  \\
& \multicolumn{1}{c|}{}                       & \cellcolor[rgb]{ .988,  .894,  .839}$\mathbf{Y}_\text{query}, \mathbf{Y}^\text{ref}_B$ &  $0.3105$   &   $0.2903$   &  $0.3219$ &  $0.2780$ \\ \cline{2-7} 
& \multicolumn{1}{c|}{\multirow{3}{*}{}}  & \cellcolor[rgb]{ 1,  .949,  .8} $\mathbf{Y}_\text{query}, \mathbf{Y}_\text{true}$ &  $6.70$   &  $6.70$    & $6.70$ & $6.70$   \\
& \multicolumn{1}{c|}{\textit{rmsd}$(\downarrow)$} & \cellcolor[rgb]{ 1,  .902,  .6} $\mathbf{Y}_\text{query}, \mathbf{Y}^\text{pred}_B$ &  $11.07$   &   $12.09$   &  $9.46$ & $10.25$   \\
& \multicolumn{1}{c|}{}  & \cellcolor[rgb]{ .988,  .894,  .839}$\mathbf{Y}_\text{query}, \mathbf{Y}^\text{ref}_B$ &  $10.27$   &  $10.71$    & $9.53$  & ${10.13}$  \\ \hline
\multirow{4}{*}{Site Quality} & \multicolumn{1}{c|}{\multirow{2}{*}{\textit{TM}$(\uparrow)$}} & \cellcolor[rgb]{ .851,  .882, .949} $\mathbf{Y}_\text{true}, \mathbf{Y}^\text{pred}_B$  &  ${0.4652}$   &  $0.4421$    & ${0.4323}$  & ${0.3889}$  \\
& \multicolumn{1}{c|}{}& \cellcolor[rgb]{ .706,  .776,  .906} $\mathbf{Y}_\text{true}, \mathbf{Y}^\text{ref}_B$ & $0.4747$   &  $0.4202$    & $0.4396$ &  $0.3725$  \\ \cline{2-7} 
& \multicolumn{1}{c|}{\multirow{2}{*}{\textit{rmsd}$(\downarrow)$}}& \cellcolor[rgb]{ .851,  .882, .949}$\mathbf{Y}_\text{true}, \mathbf{Y}^\text{pred}_B$ &  $9.77$   &  $11.76$    &   $8.95$ & $10.49$ \\
& \multicolumn{1}{c|}{}& \cellcolor[rgb]{ .706,  .776,  .906}$\mathbf{Y}_\text{true}, \mathbf{Y}^\text{ref}_B$ &  $8.20$   &  $8.92$    & $8.15$  & ${8.91}$  \\ 
\hline
\hline
\end{tabular}
}
    \caption{\textit{dockQ}, \textit{TM}, and \textit{rmsd} for evaluation of PPIs and binding sites of \textbf{Top1, Top10 hit} predicted by PPIretrieval in comparison with ground-truth structures on cross-datasets over three runs. The database for each test set comprises surface representations from the training and validation sets of each respective dataset.}
  \label{tab:pdb.vs.ppbs.crossdataset}%
\end{table}

\begin{figure}[ht!]
\subfigure[PDB Model Cross-Validated on PPBS Test Set]{{\includegraphics[width=0.5\columnwidth]{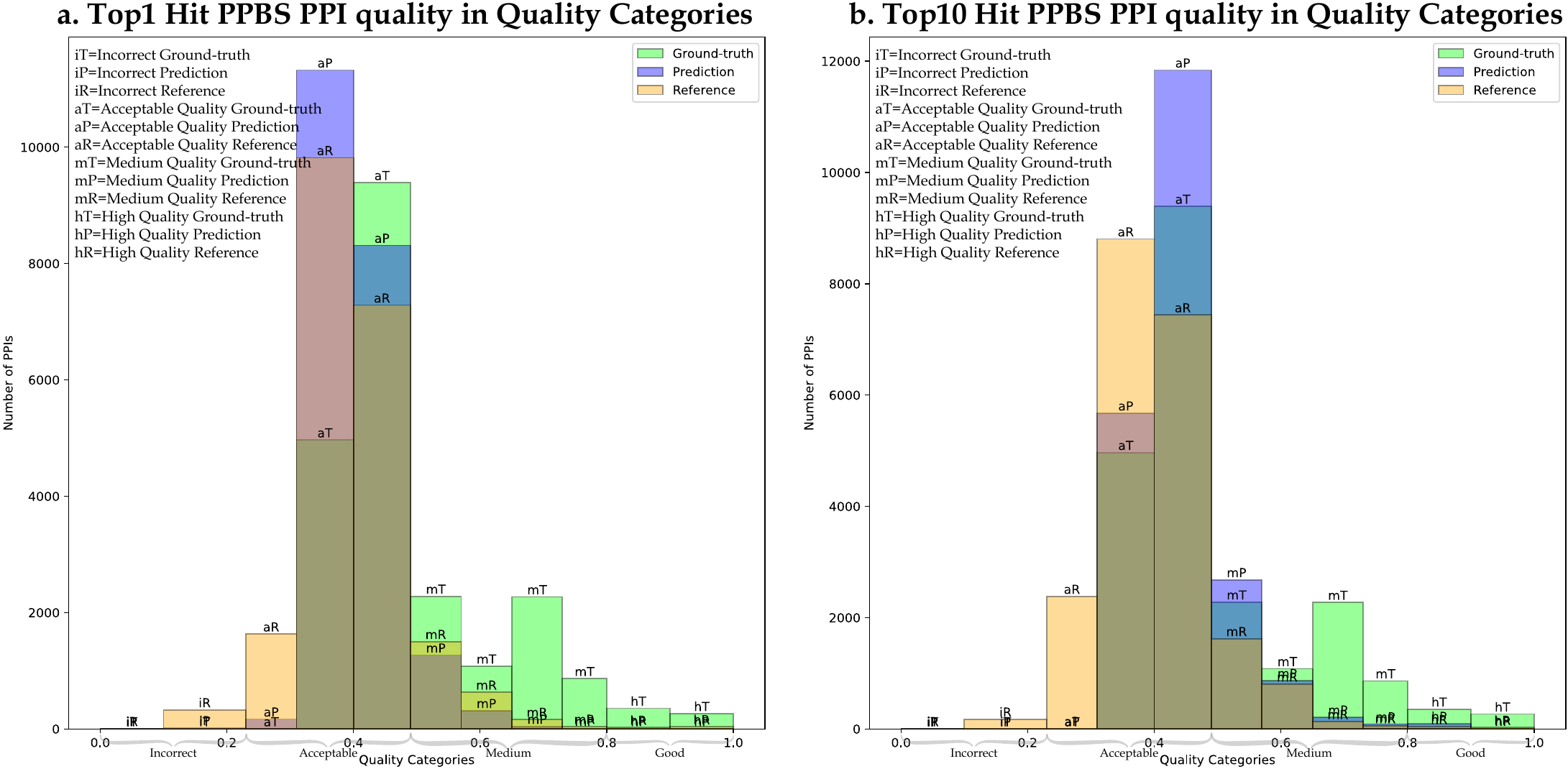}}}
\subfigure[PPBS Model Validated on PPBS Test Set]{{\includegraphics[width=0.5\columnwidth]{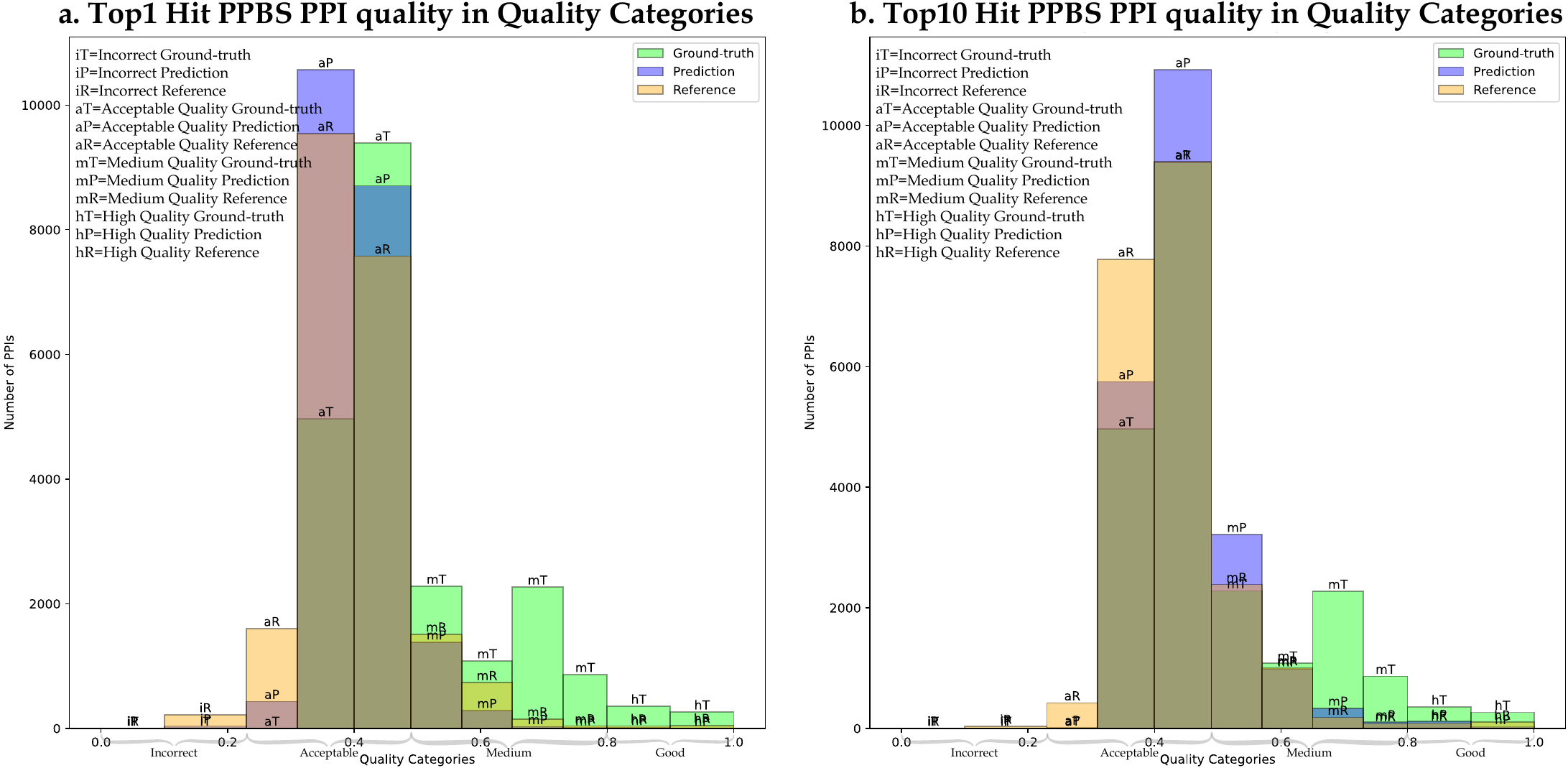}}}
\caption{Comparison of PPI qualities in the test set of PPBS dataset, considering ground-truth, predicted, and reference PPIs, evaluated using the \textit{dockQ} score at \textbf{Top1, Top10 hit}. (a) presents results from the model trained on the PDB training set and cross-validated on the PPBS test set. (b) presents results from the model trained on the PPBS training set and validated on the PPBS test set.}
\label{fig:pdb.ppbs.crossdataset}
\end{figure}

In addition to the results presented in Table~\ref{tab:pdb.vs.ppbs.crossdataset}, we also provide a visualization of the quality distribution of predicted PPIs by PPIretrieval for the cross-dataset validation in Fig.~\ref{fig:pdb.ppbs.crossdataset}. Fig.~\ref{fig:pdb.ppbs.crossdataset}(a) displays the results of the model trained on the PDB training set and then cross-validated on the PPBS test set, while Fig.~\ref{fig:pdb.ppbs.crossdataset}(b) shows the results of the model trained on the PPBS training set and then validated on the PPBS test set. With \textbf{Top1 hit}, it is noticeable that both models predict PPIs of approximately the same quality, as evidenced by comparing the quality distributions. This visual representation demonstrates that PPIretrieval possesses the capability to generalize to unseen protein complexes and accurately predict their interactions.

\section{Additional Experiments on Binding Interface Prediction}
\label{appen:baseline.models}
We show additional results on the binding interface prediction task. 

We evaluate the models on PDB and PPBS datasets (details reported in Sec.~\ref{sec:experiment}).
For baseline comparisons, we train MaSIF-search \citep{gainza2020deciphering} and dMaSIF-search\footnote{The MaSIF and dMaSIF models were originally designed without prediction capabilities for binding interfaces of protein complexes. We have modified their architectures to enable the models to make predictions for binding interfaces.} \citep{sverrisson2021fast} to predict the binding interfaces of receptor and ligand proteins in a complex. The comparisons are summarized and highlighted in Tab.~\ref{tab:ppisearch.baseline}.
MaSIF \citep{gainza2020deciphering} and dMaSIF \citep{sverrisson2021fast} offer model variants, namely MaSIF-search and dMaSIF-search, for predicting interactions between protein complexes. To enable these models to make predictions for the binding interfaces of protein complexes, we extract their final embeddings, denoted as $\mathbf{F}_R \in \mathbb{R}^{M_R \times d}$ and $\mathbf{F}_L \in \mathbb{R}^{M_L \times d}$, just before the output layer. 
Employing the strategy outlined in Sec.~\ref{sec:interactive.decoder} to predict the surface-level binding interface, we employ a MLP with sigmoid function directly on these embeddings, resulting in $\hat{\mathbf{Y}}^{\text{surf}}_R \leftarrow \sigma(\text{MLP}(\mathbf{F}_R))\in [0,1]^{M_R\times 1}, \hat{\mathbf{Y}}^{\text{surf}}_L \leftarrow \sigma(\text{MLP}(\mathbf{F}_L))\in [0,1]^{M_L\times 1}$.
We then compute the residue-level binding interfaces, denoted as $\hat{\mathbf{Y}}^{\text{res}}_R \in \{0, 1\}^{N_R \times 1}, \hat{\mathbf{Y}}^{\text{res}}_L \in \{0, 1\}^{N_L \times 1}$ from $\hat{\mathbf{Y}}^{\text{surf}}_R, \hat{\mathbf{Y}}^{\text{surf}}_L$, respectively. For each residue $i$ in $R$, we define a region with a fixed radius of $r=10\mathring{\text{A}}$ and collect a set of surface points within this region, each with a binding interface $\hat{\mathbf{y}}^{\text{surf}}_j$ and embedding $\mathbf{f}_j$. The residue $i$ is considered part of the binding interface if the majority of surface points in the region are labeled as part of the binding interface, i.e., $\hat{\mathbf{y}}^{\text{res}}_i = 1$ if $\text{Mean}(\sum_j \hat{\mathbf{y}}^{\text{surf}}_j) > 0.5$; otherwise $\hat{\mathbf{y}}^{\text{res}}_i = 0$. The residue-level binding interface $\hat{\mathbf{Y}}^{\text{res}}_L $ for $L$ is computed by the same method. In terms of training strategy, these baseline models are directly optimized through the binding interface optimization in Sec.~\ref{sec:binding.intergace.optimization}.

\begin{table}[htbp!]
\footnotesize
  \centering
  \resizebox{0.5\columnwidth}{!}{%
    \begin{tabular}{c|c|c|c|c|c|c}
    \hline
    \hline
    \rowcolor[rgb]{ .9,  .9,  .9} {\textbf{Dataset}} & \multicolumn{2}{c|}{\cellcolor[rgb]{ .851,  .882,  .949}PDB} & \multicolumn{2}{c}{\cellcolor[rgb]{ .988,  .894,  .839}PPBS} \\
    \hline
    \rowcolor[rgb]{ .851,  .851,  .851} {\textbf{Model}} & \textbf{Acc}$(\uparrow)$ & \textbf{ROC}$(\uparrow)$ & \textbf{Acc}$(\uparrow)$ & \textbf{ROC}$(\uparrow)$ \\
    \hline
    MaSIF-search & $22.93$    & $20.10$ & $20.45$     & $19.79$ \\
    \hline
    dMaSIF-search & $20.86$     & $20.74$      &  $22.35$     & $21.08$ \\
    \hline
    PPIretrieval  & $\mathbf{92.76}$     & $\mathbf{92.61}$    & $\mathbf{93.55}$     & $\mathbf{94.98}$ \\
    \hline
    \hline
    \end{tabular}%
    }
    \caption{Accuracy and ROC of PPIretrieval in comparison with MaSIF-search and dMaSIF-search on datasets over 5 runs.}
  \label{tab:ppisearch.baseline}%
\end{table}%

In Table~\ref{tab:ppisearch.baseline}, we observe that PPIretrieval surpasses MaSIF-search and dMaSIF-search, achieving improvements of $69.83\%, 71.90\%$ in accuracy, and $72.51\%, 71.87\%$ in ROC on the PDB dataset. Additionally, PPIretrieval outperforms MaSIF-search and dMaSIF-search by $73.10\%, 71.20\%$ in accuracy, and $75.19\%, 73.90\%$ in ROC on the PPBS dataset.

\paragraph{Abalation Study}
In Tab.~\ref{tab:ppisearch.ablation}, we conduct an ablation study to examine the effectiveness of our training objectives for the PPIretrieval model, which includes the \textit{lock-and-key} goal $\mathcal{L}_{\text{match}}$, the contrastive goal $\mathcal{L}_{\text{contra}}$, and the binding interface goal $\mathcal{L}_{\text{bind}}$.

\begin{table}[htbp!]
\footnotesize
  \centering
  \resizebox{0.5\columnwidth}{!}{%
    \begin{tabular}{c|c|c|c|c|c}
    \hline
    \hline
    \rowcolor[rgb]{ .8,  .8,  .8}\textbf{Dataset} & \cellcolor[rgb]{ .851,  .851,  .851}$\mathcal{L}_\text{match}$ & \cellcolor[rgb]{ .851,  .851,  .851}$\mathcal{L}_\text{contra}$ & \cellcolor[rgb]{ .851,  .851,  .851}$\mathcal{L}_\text{bind}$ & \cellcolor[rgb]{ .851,  .851,  .851}\textbf{Acc}$(\uparrow)$ & \cellcolor[rgb]{ .851,  .851,  .851}\textbf{ROC}$(\uparrow)$  \\
    \hline
    \cellcolor[rgb]{ .851,  .882,  .949}PDB   &       &       & $\checkmark$     &      $91.10 $ & $90.88 $ \\
    \cellcolor[rgb]{ .851,  .882,  .949}PDB   &       & $\checkmark$     & $\checkmark$     &  $92.11 $     &  $91.13 $ \\
    \cellcolor[rgb]{ .851,  .882,  .949}PDB   & $\checkmark$     &       & $\checkmark$     &  $90.86$     & $91.95$ \\
    \cellcolor[rgb]{ .851,  .882,  .949}PDB   & $\checkmark$     & $\checkmark$     & $\checkmark$     &$ \mathbf{92.76 }$ & $ \mathbf{92.61}$ \\
    \hline
    \hline
    \end{tabular}%
    }
    \caption{Ablation study of PPIretrieval optimization. A checkmark $(\checkmark)$ indicates that an objective is used to optimize the model.}
    \vspace{-0.2cm}
  \label{tab:ppisearch.ablation}%
\end{table}%
We observe that, overall, PPIretrieval performs better in terms of both accuracy and ROC when all three optimization objectives are combined. This suggests that PPIretrieval successfully learns the \textit{lock-and-key} structure between the receptor and ligand in a protein complex under our training strategy.

\begin{figure}[ht!]
\centering
{\includegraphics[width=.5\columnwidth]{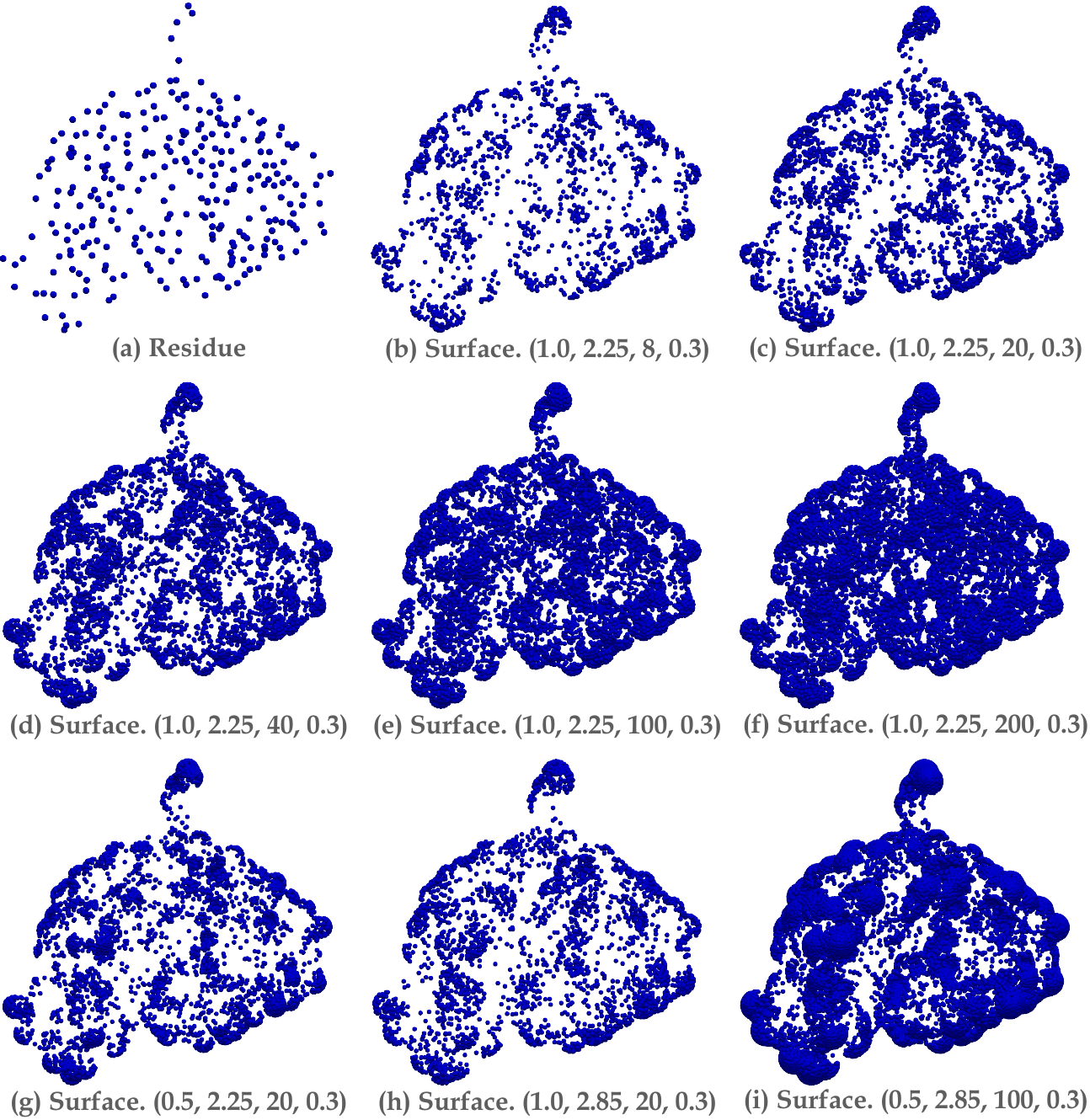}}
\vspace{-0.3cm}
  \caption{Visualization of dMaSIF parameters for surface sampling. The bracket includes (Resolution, Distance, \#Sample, Variance).
  \label{fig:dMaSIF.sampling}
}
\end{figure}

In Fig.~\ref{fig:dMaSIF.sampling}, we visualize protein surface sampling with different dMaSIF parameters. 
We can observe that lower sampling resolution, higher sampling distance, and a greater number of sampling points contribute to a more accurate approximation of the protein surface manifold. However, due to limited computational resources, we can only perform sampling and training shown in Fig.~\ref{fig:dMaSIF.sampling}(c). Training PPIretrieval on more precise surface sampling with more computing resources is a potential direction for exploration.

In Tab.~\ref{tab:dMaSIF.ablation}, we present the performance of PPIretrieval with different dMaSIF sampling parameters. 
\begin{table}[htbp!]
\footnotesize
  \centering
   \resizebox{0.5\columnwidth}{!}{%
    \begin{tabular}{c|c|c|c|c|c|c|c}
    \hline
    \hline
    \rowcolor[rgb]{ .8,  .8,  .8} \textcolor[rgb]{ 0,  0,  0}{Dataset} & \cellcolor[rgb]{ .851,  .851,  .851}Resolution & \cellcolor[rgb]{ .851,  .851,  .851}Distance & \cellcolor[rgb]{ .851,  .851,  .851}\#Sample & \cellcolor[rgb]{ .851,  .851,  .851}Variance & \cellcolor[rgb]{ .851,  .851,  .851}\#Surf Point & \cellcolor[rgb]{ .851,  .851,  .851}\textbf{Acc}$(\uparrow)$ & \cellcolor[rgb]{ .851,  .851,  .851}\textbf{ROC}$(\uparrow)$ \\
    \hline
    \cellcolor[rgb]{ .851,  .882,  .949}PDB   & 1.00     & 2.25  & 8     & 0.30   & 2175      &   $90.16$    &  $90.96$ \\
    \cellcolor[rgb]{ .851,  .882,  .949}PDB   & 1.00     & 2.25  & 12    & 0.30   &  2425   & $91.44$     & $90.99$ \\
    \cellcolor[rgb]{ .851,  .882,  .949}PDB   & 1.00     & 2.85  & 20    & 0.30   &  2632     &  $92.03$     & $91.55$ \\
    \cellcolor[rgb]{ .851,  .882,  .949}PDB   & 1.00     & 2.25  & 20    & 0.30   &   2680    & $92.76$ & $92.61$ \\
    \hline
    \hline
    \end{tabular}%
    }
    \vspace{-0.3cm}
    \caption{Ablation study of dMaSIF parameters for PPIretrieval. \textit{\#Surf Point} denotes the average number of surface points of a protein sampled by dMaSIF across the training set.}
  \label{tab:dMaSIF.ablation}%
\end{table}%
PPIretrieval exhibits improved accuracy and ROC on PDB, with more precise protein surface sampling. Once again, these results motivate further exploration for future enhancements.

\paragraph{Computational Resources}
Our models are trained on a single Nvidia 48G A40 GPU. Regarding training time, dMaSIF-search takes approximately $0.29$s to train a protein complex, while PPIretrieval takes around $0.35$s for the same task. In terms of inference time, dMaSIF-search requires about $0.10$s for a protein complex, while PPIretrieval takes approximately $0.11$s.


\end{document}